\documentclass[prd,letterpaper,twocolumn,superscriptaddress,preprintnumbers,nofootinbib]{revtex4-1}

\usepackage{slashed}
\usepackage{amsmath,amssymb}
\usepackage{bbm}
\usepackage{graphicx}
\usepackage{caption}
\usepackage{adjustbox}
\usepackage{units}
\usepackage{bbold}
\usepackage{xcolor}
\usepackage{dsfont}
\usepackage{caption}
\usepackage{subcaption}
\definecolor{naviBlue}{RGB}{0,0,128}
\usepackage[colorlinks  = true,
            linkcolor   = naviBlue,
            urlcolor    = naviBlue,
            citecolor   = naviBlue,
            anchorcolor = naviBlue]{hyperref}

\usepackage[makeroom]{cancel}

\usepackage{comment}

\usepackage{orcidlink}

\DeclareMathOperator{\tr}{Tr}

\newcommand{\beq}{\begin{equation}}
\newcommand{\eeq}{\end{equation}}
\newcommand{\bea}{\begin{eqnarray}}
\newcommand{\ena}{\end{eqnarray}}
\newcommand{\dd}{{\rm d}}

\newcommand{\eq}{\mathrm{eq}}

\usepackage[normalem]{ulem}

\begin{document}

\preprint{P3H-26-052, TTP26-022, TTK-26-19}

\title{Conversion-Driven Baryogenesis in Flavored Dark Matter Models
}

\author{Benedetta Belfatto}
\email{benedetta.belfatto@kit.edu}
\thanks{\orcidlink{0000-0002-6191-8494}}
\affiliation{\footnotesize Institut f\"ur Theoretische Teilchenphysik,
  Karlsruhe Institute of Technology, 76128~Karlsruhe, Germany\\ \looseness=-1}

\author{Monika Blanke}
\email{monika.blanke@kit.edu}
\thanks{\orcidlink{0000-0003-0413-4455}}
\affiliation{\footnotesize Institut f\"ur Theoretische Teilchenphysik,
  Karlsruhe Institute of Technology, 76128~Karlsruhe, Germany\\ \looseness=-1}
\affiliation{\footnotesize Institut f\"ur Astroteilchenphysik,
  Karlsruhe Institute of Technology, 76128~Karlsruhe, Germany\\ \looseness=-1}
  
\author{Jan Heisig}
\email{heisig@physik.rwth-aachen.de}
\thanks{\orcidlink{0000-0002-7824-0384}}
\affiliation{\footnotesize Institute for Theoretical Particle Physics and Cosmology, RWTH Aachen University, 52056 Aachen, Germany\\ \looseness=-1}

\author{Lena Rathmann}
\email{lena.rathmann@kit.edu}
\thanks{\orcidlink{0000-0003-1035-1923}}
\affiliation{\footnotesize Institut f\"ur Astroteilchenphysik,
  Karlsruhe Institute of Technology, 76128~Karlsruhe, Germany\\ \looseness=-1}
  
\author{Felix Wilsch}
\email{felix.wilsch@physik.rwth-aachen.de}
\thanks{\orcidlink{0000-0003-2409-1579}}
\affiliation{\footnotesize Institute for Theoretical Particle Physics and Cosmology, RWTH Aachen University, 52056 Aachen, Germany\\ \looseness=-1}

\begin{abstract}
The dark matter and baryon asymmetry problems remain two of the most pressing questions in fundamental physics. Considering lepton-flavored dark matter, it has recently been shown that the cogenesis of dark matter and the baryon asymmetry can be economically achieved  via  
conversion-driven freeze-out.
This mechanism leverages semi-efficient conversions to drive a departure from equilibrium while preserving independence from initial conditions through early thermalization of the dark sector.
In this work, we develop this mechanism further, providing a detailed analysis 
of the chemical-equilibrium conditions and
demonstrating that the framework can be extended to quark-philic scenarios, where the matter-antimatter asymmetry is generated resonantly 
through baryon-number-conserving $CP$-violating conversions of a mediator field into Standard Model quarks and dark matter. 
The strong QCD interactions of the colored mediator, including bound-state formation effects during freeze-out, substantially enlarge the viable parameter space and allow dark matter masses from a few hundred GeV up to the TeV scale. 
We furthermore assess the impact of thermal effects by comparing a minimal treatment with a setup that approximately accounts for thermal masses and their kinematic consequences. The resulting scenario predicts striking long-lived particle signatures with soft displaced decay products that remain only partially explored at the LHC and motivate dedicated searches at the HL-LHC.
\end{abstract}

\maketitle

\section{Introduction}
\label{sec:intro}

The origins of dark matter (DM) and the baryon asymmetry in our Universe (BAU) continue to pose two of the most compelling open questions in modern physics. Both phenomena point to the presence of physics beyond the Standard Model (SM), driving a broad range of theoretical and experimental investigations~\cite{Bertone:2004pz,Davidson:2008bu,Bodeker:2020ghk}. While a variety of mechanisms has been proposed to address these puzzles individually, a particularly appealing approach is to link them, sometimes referred to as the \emph{cogenesis} of DM and the BAU and explored in models such as sterile neutrinos~\cite{Asaka:2005pn,Canetti:2012vf} and asymmetric DM~\cite{Bento:2001rc,Berezhiani:2008zza,Petraki:2013wwa}.

In the recent work of Ref.~\cite{Heisig:2024mwr}, a novel mechanism for baryogenesis was introduced, 
which leverages
conversion-driven freeze-out~\cite{Garny:2017rxs,DAgnolo:2017dbv} of DM
-- a mechanism in which the DM abundance is driven by the partial decoupling of inelastic DM scattering instead of DM annihilations -- to generate both the observed relic density and the baryon asymmetry, without invoking a DM asymmetry. The semi-efficient conversions between DM and a mediator particle that is relatively close in mass, present in this mechanism satisfies the out-of-equilibrium Sakharov condition. By virtue of $CP$-violating contributions to the conversion processes, the mechanism can generate a sizable baryon asymmetry. Remarkably, in conversion-driven freeze-out, DM thermalizes initially, such that both the resulting abundances and asymmetries are largely insensitive to initial conditions~\cite{Garny:2017rxs} (see also \cite{Heisig:2025gdg} for the explicit effect on the asymmetry). In Ref.~\cite{Heisig:2024mwr}, the mechanism was introduced in a leptophilic scenario and hence dubbed \emph{conversion-driven leptogenesis}. 
In this work, we further develop the mechanism and
demonstrate that its generalization to quark-philic models is equally viable and provides solutions up to larger masses with interesting implications for collider searches.

Conversion-driven baryogenesis shares similarities with Dirac leptogenesis~\cite{Dick:1999je}. There, the fact that electroweak sphalerons act only on left-handed particles allows the lepton asymmetry to be generated by partly storing lepton number in a decoupled right-handed sector, without requiring lepton-number-violating interactions. In the original realization with Dirac neutrino masses, $CP$-violating out-of-equilibrium decays generate 
equal but opposite sign
lepton asymmetries in the SM sector and
in right-handed neutrinos,
whose tiny Yukawa couplings keep them decoupled until (and in fact beyond) sphaleron freeze-out. Related mechanisms have been proposed in which $B$-conserving out-of-equilibrium decays of self-conjugate particles populate two sectors
carrying baryon number 
while only one is affected by
electroweak sphalerons~\cite{AristizabalSierra:2013lyx}; see also Ref.~\cite{Cui:2011ab} for a realization based on DM annihilation.

\begin{figure*}[t]    
  \centering
\vspace{2ex}
\includegraphics[width=0.8\textwidth, trim= {0.0cm 0.0cm 0.0cm 0.0cm}, clip]{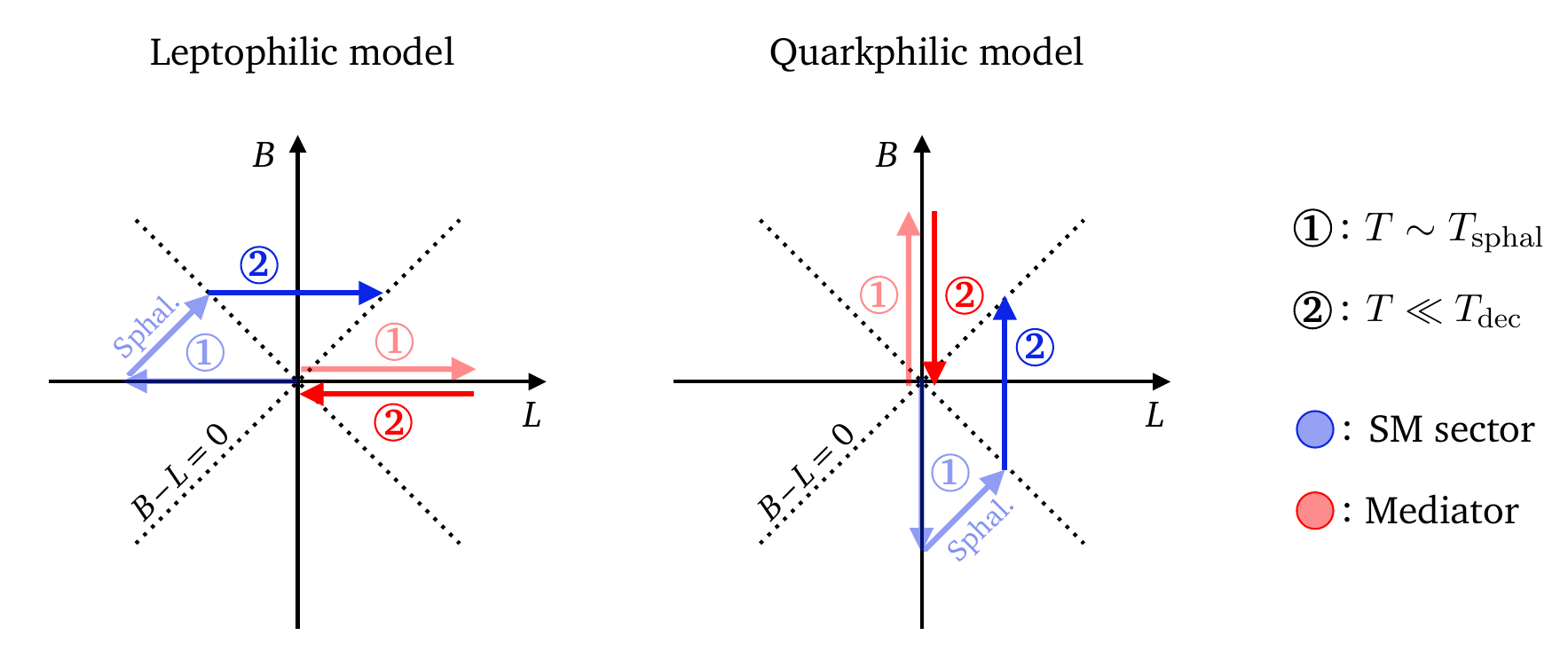} 
\vspace{-1ex}
  \caption{
Schematic diagram showing two snapshots of the evolution of $B$ and $L$ in the leptophilic and quarkphilic model.
}
  \label{fig:BLdiag}
\end{figure*}

As we detail in this paper, in
conversion-driven baryogenesis, 
a baryon asymmetry can likewise be generated from $B$ and $L$ conserving 
$CP$-violating conversions of two sectors with different exposure to electroweak sphalerons.
The mediator couples to a right-handed SM fermion multiplet and carries the corresponding baryon or lepton quantum number, so that $B$ and $L$ remain perturbatively conserved.
While gauge interactions keep the mediator in thermal equilibrium with the SM plasma, 
the DM departs from equilibrium before sphaleron decoupling due to its small interactions with the other fields.
SM Yukawa interactions then
communicate the asymmetry generated in the right-handed fermion sector 
to the sphaleron-active left-handed degrees of freedom.
Electroweak sphalerons thereby partially convert the SM-sector asymmetry into a baryon asymmetry. 
Although the mediator remains in thermal equilibrium with the SM plasma, its equilibrium interactions 
do not communicate its
$B-L$ asymmetry to the sphaleron-active left-handed sector, and the compensating mediator asymmetry is transferred back to the SM only later, through mediator decays.
Unlike in Dirac leptogenesis, the relevant new-physics particle carrying baryon/lepton number
is therefore not a thermally decoupled sterile neutrino with a persistent abundance, but a mediator that remains coupled to the thermal plasma and eventually decays.

The mechanism is illustrated in the schematic $B$–$L$ diagrams of Fig.~\ref{fig:BLdiag}, displaying the leptophilic (left panel) and quark-philic (right panel) cases. The figure shows two snapshots of the evolution of lepton and baryon number in the SM (blue) and new-physics sector (mediator; red). The pale colors illustrate the effect of asymmetry generation up to the point of sphaleron decoupling, $T \sim 130\,$GeV, which is accompanied by $B+L$ washout processes respecting $B-L$. The solid colors show the effect of the mediator decay (well after sphaleron decoupling). 
With the decay of the mediator particle, its asymmetry is transferred to the corresponding SM fermions and cancels the respective counterpart of the mediator asymmetry in the SM sector. Since sphaleron processes have meanwhile reshaped the SM asymmetries, a net baryon asymmetry remains in both cases.
The details depend on the underlying chemical potentials involved.

In the following, we will first introduce the class of models used in Sec.~\ref{sec:model}. We will then derive the relation between the chemical potentials and asymmetries in the number densities in Sec.~\ref{sec:bgenmech}. This allows us to formulate the Boltzmann equations and demonstrate the viability of the mechanism for quark-philic models in Sec.~\ref{sec:BMEs} before discussing the allowed parameter space in Sec.~\ref{sec:viableparam}.

\section{Flavored dark matter model}
\label{sec:model}

Conversion-driven freeze-out can be realized in minimal $t$-channel mediator DM models, yet to simultaneously generate a baryon asymmetry $CPT$ invariance requires multiple decay modes of the mediator to be present for a non-vanishing $CP$ asymmetry. This condition is met in flavored DM models (see \cite{Kile:2011mn,Batell:2011tc,Agrawal:2011ze} for early accounts) where 
we supplement the SM by a gauge singlet DM multiplet and a $t$-channel mediator with the same gauge quantum numbers as the SM fermions it couples to. The new particles are assumed to be odd under a new $\mathbb{Z}_2$~symmetry while SM particles are even. A recent overview of different flavored DM models can be found in~\cite{Belfatto:2025ids}. For definiteness, we consider a Majorana fermion DM multiplet~$\chi_i$ and a scalar mediator~$\phi$ coupling to right-handed SM fermions,~$f_\alpha$. 

The described particle content allows for the following renormalizable Lagrangian:
\begin{align}
    \mathcal{L} \,\supset\, & \frac{1}{2}  \bar{\chi_i}  \,\slash\!\!\!\!\!\;\partial \,\chi_i-\frac{M_\chi^{ij}}{2} \bar{\chi_i} \chi_j+ (D_{\mu}\phi)^\dagger \, D^{\mu}\phi -m_\phi^2 |\phi|^2 \nonumber \\
    &- \
  (\lambda_{\alpha i} \phi  \,\bar f_{\text{R},\alpha} \chi_i  \ + \ \text{h.c.} ) -\lambda_H H^{\dag} H \phi^{\dag} \phi\,,
\end{align}
where $m_\phi$ is the mediator mass, $M_\chi=\text{diag}(m_{\chi_1},\dots, m_{\chi_n})$ is the diagonal Majorana mass matrix with mass eigenstates~$\chi_i$, $\lambda$~a complex Yukawa matrix, $\lambda_H$~the Higgs-portal coupling, and 
$D_\mu$~the covariant derivative. For later convenience, we introduce the overall Yukawa coupling strength 
\begin{equation}\label{eq:lambda-bar}
\bar{\lambda}\equiv \sqrt{\tr(\lambda^\dagger\lambda)}\,,
\end{equation}
that enters the total mediator decay width, $\sum_{i,\alpha}\Gamma_{\phi\rightarrow \chi_i f_\alpha} \propto \bar{\lambda}^2$.
\smallskip

Note that the mass matrix~$M_\chi$ is expected to receive radiative contributions from the multiplet's interaction with the SM via~$\lambda$. In fact, in dark minimal flavor violation (DMFV)~\cite{Agrawal:2014aoa} one assumes that  those corrections provide the only sources of mass splittings among the multiplet states. However, working in the mass eigenbasis it becomes apparent that in DMFV, the $CP$ asymmetry resulting from the interactions $\lambda$ vanishes~\cite{Heisig:2024mwr, Dev:2015wpa}. 
Hence, while viable realizations of conversion-driven baryogenesis in flavored DM models favor a DM mass splitting scale in the ballpark of the radiative corrections, it requires contributions beyond DMFV\@.

In order to realize conversion-driven freeze-out in flavored DM models, DM annihilations and DM-mediator coannihilations need to be sufficiently suppressed, the conversion processes need to become semi-efficient during freeze-out, and mediator-pair annihilations need to efficiently deplete the dark sector. This typically requires a very small coupling ($\bar\lambda$ of the order of $10^{-6}$) together with a relatively small DM-mediator mass splitting up to a few tens of GeV~\cite{Garny:2017rxs,Acaroglu:2023phy}. The small coupling suppresses DM annihilations and coannihilations while allowing conversions to become inefficient around freeze-out. A small mass splitting ensures that the mediator population is not too Boltzmann suppressed, such that its annihilations can continue to deplete the total dark-sector abundance through conversions. The transition to the canonical freeze-out regime for larger mass splittings is very sharp, with the value of the coupling $\lambda$ required to obtain the observed relic density increasing  by several orders of magnitude for only a small increase in the mass splitting. We refer to this sharp transition as the conversion-driven freeze-out (CDFO) boundary, which will be visualized in Sec.~\ref{sec:viableparam}.

\section{Number densities and chemical potentials}
\label{sec:bgenmech}

The baryon number of the Universe is the result of the asymmetries between particles and antiparticles number densities.
These asymmetries are related to chemical potentials, $\mu_X$, of each species $X$.
At first order in $|\mu_X /T|\ll 1$ they read:
\begin{align}
&n_{\Delta X} \equiv\  n_X-n_{\bar{X}}= \frac{g_X T^3}{6} \frac{\mu_X}{T}\, k_X
\end{align}
where $n_X$ and $n_{\bar{X}}$ are the number densities of the particle $X$ and its $CP$-conjugate state, $\bar X$, respectively, 
and $g_X$ its degrees of freedom. Furthermore, $T$~denotes the temperature and 
\begin{align}
k_X = 
\!\left\{
\begin{array}{l}
    1 \hspace{.5cm} \text{relativistic fermions}
    \\[2mm]
   2 \hspace{.5cm} \text{relativistic bosons}
     \\[2mm]
 \frac{6}{\pi^2}\big(\frac{m_X}{T}\big)^{\!2} K_2\big(\frac{m_X}{T}\big)\!\underset{m_X\gg T}{\approx} \!\! \frac{ 6\, e^{-m_X/T} m_X^{3/2}}{ \sqrt{2} \pi^{3/2}\, T^{3/2}} \hspace{.35cm} \text{non rel.}
\end{array}\right.
\end{align}
where $K_2(x)$ is the modified Bessel function of the second kind of order~2. 

The evolution of particle number asymmetries can be tracked by Boltzmann equations. However, asymmetries
of different species are not independent of each other, since fast interactions ($\Gamma \gg H $) and conservation laws establish equilibrium conditions
on chemical potentials. Slow processes instead need to be taken into account in evolution equations.

In particular, the following conditions are enforced by fast SM processes when the baryogenesis happens above the electroweak phase-transition temperature (symmetric phase) \cite{Harvey:1990qw}.
Assuming chemical potentials are equal between families 
(rapid flavor-changing interactions, $\mu_{\ell i}=\mu_\ell$ for lepton doublets, $\mu_{ei}=\mu_e$ for lepton singlets, $\mu_{q i}=\mu_q$ for quark doublets, $\mu_{ui}=\mu_u$ and $\mu_{di}=\mu_d$ for quark singlets),
thermal equilibrium of Yukawa interactions implies
\begin{align}
& \mu_{q}+\mu_H -\mu_{u}=0 \, , \nonumber \\
& \mu_{q}-\mu_H -\mu_{d}=0 \, , \nonumber \\
& \mu_{\ell }-\mu_H -\mu_{e}=0
 \label{eq:yuk-eq}
\end{align}
with $\mu_H$ the chemical potential of the Higgs doublet. Sphaleron interactions in equilibrium enforce the relation\footnote{$SU(3)$ QCD instanton processes bring the relation $\sum_i (2\mu_{qi}-\mu_{ui}-\mu_{di})=0$ for SM quarks \cite{Mohapatra:1991bz}. This condition is automatically satisfied if the quark Yukawa interactions are in equilibrium.}
\begin{align}
&  (3\mu_{q}+\mu_{\ell })=0
 \label{eq:sph-eq}
\end{align}
The condition of hypercharge neutrality of the universe ($y=0$) gives:
\begin{align}
& y=3\,( \mu_q +2 \mu_u -\mu_d -\mu_\ell -\mu_e)+2\mu_H +y_\phi \, g_\phi  k_\phi \mu_\phi =0 \, ,
 \label{eq:hyp}
\end{align}
where $y_X$ is the hypercharge of particle species $X$.

Baryon and lepton number densities $n_{\Delta B}=BT^2/6$, $n_{\Delta L}=LT^2/6$ can be written in terms of chemical potentials.
 From Eqs. \eqref{eq:yuk-eq}--\eqref{eq:hyp} we have
 \begin{align}
  B &= 3 \, \big( 2\,\mu_q+\mu_u+\mu_d\big)+N^\phi_B\, g_\phi k_\phi\mu_\phi = 
\nonumber \\
& =12\mu_q+N^\phi_B g_\phi k_\phi\mu_\phi \nonumber \\
 L & = 3\, \big( 2\, \mu_\ell+\mu_e \big)+N^\phi_L\, g_\phi k_\phi\mu_\phi =
\nonumber \\ &
= -\frac{153}{7}\mu_q + \Big( \frac{3}{14}y_\phi +N_L^\phi \Big)\, g_\phi k_\phi\mu_\phi 
\,,
\end{align}
where $N^\phi_{B\,(L)}$ denotes the baryon (lepton) number assigned to the mediator~$\phi$.
There is one additional condition coming from a conserved quantum number, that is, the conservation of $B-L$.
After imposing 
$B-L=(B-L)_{\text{initial}}=0$,
baryon and lepton number asymmetries can be expressed in terms of only one chemical potential (e.g. $\mu_\phi$).
Then, we only need to solve the equation for the number density of $\phi,\bar{\phi}$.

Below the temperature of electroweak phase transition
SM particles acquire masses, while 
the mass term of extra scalar particles is modified by the 
contribution induced by the Higgs.
Rapid electroweak interactions enforce conditions among the particle components of the same weak multiplet,
while the chemical potential of the neutral component ($\mu_0$) of the Higgs boson vanishes, 
and the chemical potential of the charged gauge bosons $\mu_W$ should be taken into account (see ref. \cite{Harvey:1990qw}).
Assuming chemical potentials are equal between families
($\mu_{u_{L(R)}}$ for all left(right)-handed up quarks, $\mu_{d_{L(R)}}$ for left(right)-handed down quarks, $\mu_{e_{L(R)}}$ for charged leptons, $\mu_{\nu_L}$ for neutrinos),
the following conditions are given by the equilibrium of weak interactions 
\begin{align}
& 
\mu_{u_L}=\mu_{d_L}-\mu_W \,, \quad \mu_{e_L}=\mu_{\nu_L}+\mu_W \,,
 \label{eq:weak-eq}
\end{align}
Yukawa interactions 
\begin{align}
& \mu_{u_L}+\mu_0 -\mu_{u_R}=0 \, , \qquad  \mu_{d_L}-\mu_0 -\mu_{d_R}=0 \, , \nonumber \\ &
 \mu_{e_L }-\mu_0 -\mu_{e_R}=0 \,,
 \label{eq:yuk-broken}
\end{align}
and rapid sphaleron interactions 
\begin{align}
&  3(\mu_{u_L}+2\,\mu_{d_L })+3\,\mu_{\nu_L}=0 \,.
 \label{eq:sph-broken}
\end{align}
The condition of charge neutrality of the universe reads:
\begin{align}
 Q =& \, (4+2k_{t})\,( \mu_{u_L} + \mu_{u_R})-3( \mu_{d_L}+\mu_{d_R} )
+\nonumber \\ &
-3(\mu_{e_L} +\mu_{e_R})-6\mu_W
+Q_\phi g_\phi k_\phi\mu_\phi=0
 \label{eq:charge}
\end{align}
where the extra particle $\phi$ has charge $Q_\phi$ and we took into account possible effects of the top mass.
After applying the conditions for fast interactions and charge neutrality, 
baryon and lepton asymmetries are given by
  \begin{align}
  B 
 & = \frac{4}{3}\big(16+10k_t+k_t^2 \big)\, \mu_{u_L}+
 \nonumber \\ & +
 \Big(\frac{1}{3}\big(2+k_t \big) Q_\phi+N^\phi_B\Big)\, g_\phi k_\phi\mu_\phi \,,
 \nonumber \\
 L 
 &=-75\, \mu_{u_L} + \Big( N_L^\phi -2Q_\phi \Big)\, g_\phi k_\phi\mu_\phi \,.
 \end{align}
The conservation of $B-L$ gives the additional condition which ensures that the baryon asymmetry 
is determined by only one chemical potential.
 
For instance, we can have the following scenarios, assuming the condition $B-L=0$:
\begin{enumerate}
    \item[\emph{(a)}]
    New species $\phi$ coupling to right-handed leptons, $y_\phi=-1$, $g_\phi=1$, $N_L^\phi=1$, $N_B^\phi=0$:
the baryon number can be written in terms of the evolution of the field $\phi$, 
in the symmetric (SP) or broken (BP) phase, respectively, as:\footnote{Note that this result differs from the one used in \cite{Heisig:2024mwr}.}
    \begin{align}
  B &=\, k_\phi\mu_\phi\, \begin{cases}
      \frac{22}{79}\quad \text{(SP)}\\
      \frac{11}{37}\quad \text{(BP)}
  \end{cases}\, 
 \end{align}
Then, the number density of the baryon asymmetry is given by
 \begin{align}
  \qquad n_{\Delta B} = \frac{T^2}{6}\, B\, = \big( n_{\phi}-n_{\bar{\phi}} \big)_{T=T_\text{Sphal}}
  \begin{cases}
      \frac{22}{79}\, \quad \text{(SP)} \\
      \frac{11}{37}
  \quad \text{(BP)} 
  \end{cases}.
 \end{align}

\item[\emph{(b)}] 
New fields coupling to right-handed up-type quarks ($y_\phi=\frac{2}{3}$, $g_\phi=3$, $N_B^\phi=\frac{1}{3}$, $N_L^\phi=0$): we would have the baryon number
    \begin{align}
  B &=\, k_\phi\mu_\phi\, \begin{cases}
      \frac{63}{79}\quad \text{(SP)}\\
      \frac{27}{37}\quad \text{(BP)}
  \end{cases}\, 
 \end{align}
and a baryon asymmetry
 \begin{align}
 \;\; n_{\Delta B}  \, = \; 
  \frac{1}{3}\big( n_{\phi}-n_{\bar{\phi}} \big)_{T=T_\text{Sphal}}
\begin{cases}
    \frac{63}{79}\, \quad \text{(SP)}\\
    \frac{27}{37}\,
  \quad \text{(BP)}
\end{cases}  \!\!.
 \end{align}

\item[\emph{(c)}]
For new fields coupling to right-handed down-type quarks ($y_\phi=-\frac{1}{3}$, $g_\phi=3$, $N_B^\phi=\frac{1}{3}$, $N_L^\phi=0$) the baryon asymmetry reads: 
  \begin{align}
 \;\; n_{\Delta B}  \, = \; 
  \frac{1}{3}\big( n_{\phi}-n_{\bar{\phi}} \big)_{T=T_\text{Sphal}}
\begin{cases}
    \frac{45}{79}\, \quad \text{(SP)}\\
    \frac{24}{37}\,
  \quad \text{(BP)}
\end{cases}  \!\!.
\label{eq:nb_down}
 \end{align}

\end{enumerate}

The asymmetries in the SM sector derive from the 
conditions of equilibrium of rapid interactions and the conservation of charges.
For instance, (assuming that the involved SM particles are relativistic, $k_f=1$) one has in the scenarios with new fields coupling to
\begin{enumerate}
\item[\emph{(a)}] right-handed leptons:
\begin{equation}
    n_{e^-}-n_{e^+}=-\,(n_\phi-n_{\bar{\phi}})\, n_F
\begin{cases}
    \frac{8}{79} \:\,\quad \text{(SP)}\\ 
    \frac{19}{222} \quad \text{(BP)}
\end{cases}\!\!,
\label{eq:Deltae}
\end{equation}
\item[\emph{(b)}] right-handed up-type quarks:
\begin{equation}
n_{u}-n_{\bar{u}}=-\,(n_\phi-n_{\bar{\phi}})\, n_F
\begin{cases}
    \frac{31}{237} \quad \text{(SP)}\\ 
    \frac{7}{111} \quad \text{(BP)}
\end{cases}\!\!,
\label{eq:Deltau}
\end{equation}
\item[\emph{(c)}] right-handed down-type quarks:
\begin{equation}
n_{d}-n_{\bar{d}}=-\,(n_\phi-n_{\bar{\phi}})\, n_F
\begin{cases}
    \frac{40}{237} \quad \text{(SP)}\\ 
    \frac{17}{222} \quad \text{(BP)}
\end{cases}\!\!,
\label{eq:Deltad}
\end{equation}
\end{enumerate}
with $n_F$ the number of families coupling to $\phi$.

Note that in the numerical analysis below, we trace the generated asymmetries down to the sphaleron decoupling temperature $T_\text{sph} \simeq 130\,$GeV while using the symmetric-phase relations. Since this temperature lies close to the electroweak crossover, a treatment of the transition region would require a more refined description of the thermal plasma. However, the conversion factors in the symmetric and broken phases differ only mildly, such that this choice has no significant impact on our results.

\section{Boltzmann equations}
\label{sec:BMEs}

Introducing $Y_X=n_X/s$, with the entropy density~$s$, $x=M_{\chi_1}/T$, and $Z=1/(1+\frac{1}{3}\frac{\dd\ln g_{*s}}{\dd \ln T})$ we can formulate the set of Boltzmann equations as:
\begin{align}
 & x \mathcal{H}Z \,\frac{\dd (Y_{\phi}+Y_{\bar{\phi}})}{\dd z} =
- \sum_{i,\alpha} \Bigg[ \langle\Gamma_{\phi\rightarrow \chi_i f_\alpha} \rangle Y_\phi + 
   \langle\Gamma_{\bar{\phi}\rightarrow \chi_i \bar{f}_\alpha} \rangle Y_{\bar{\phi}}
       \nonumber \\ &  \qquad 
  -Y_\phi^\text{eq}
\frac{Y_{\chi_i}}{Y_{\chi_i}^\text{eq}} \left( 
\frac{Y_{f_\alpha}}{Y_{f_\alpha}^\text{eq}}\, \langle\Gamma_{\bar{\phi}\rightarrow \chi_i \bar{f}_\alpha} \rangle  +\frac{Y_{\bar{f}_\alpha}}{Y_{f_\alpha}^\text{eq}}\,\langle\Gamma_{\phi\rightarrow \chi_i f_\alpha} \rangle \right) 
\Bigg]\nonumber
\\ &
\qquad -2\langle\sigma v\rangle\, s\,\big( Y_{\phi}Y_{\bar{\phi}} - (Y_\phi^\text{eq})^2\big)
\,, \label{eq:BME11}
\end{align}
%
\begin{align}
   & x \mathcal{H}Z \,\frac{\dd (Y_{\phi}-Y_{\bar{\phi}})}{\dd x} 
   =
  -\langle\Gamma_\phi \rangle \,  \big(Y_{\phi}-Y_{\bar{\phi}}\big) 
     \nonumber \\ & \hspace{.4cm}
+ \, Y_\phi^\text{eq}  \, 
\sum_{i,\alpha} \frac{Y_{\chi_i}}{Y_{\chi_i}^\text{eq}} 
\bigg[ \frac{Y_{f_\alpha}+Y_{\bar{f}_\alpha}}{2\,Y_{f_\alpha}^\text{eq}}\,
\Big( 
 \langle\Gamma_{\bar{\phi}\rightarrow \chi_i \bar{f}_\alpha} \rangle  -\langle\Gamma_{\phi\rightarrow \chi_i f_\alpha} \rangle \Big) 
   \nonumber \\ & \hspace{.4cm} 
   +
\frac{Y_{f_\alpha}-Y_{\bar{f}_\alpha}}{2\,Y_{f_\alpha}^\text{eq}}\,
\Big( 
 \langle\Gamma_{\bar{\phi}\rightarrow \chi_i \bar{f}_\alpha} \rangle  +\langle\Gamma_{\phi\rightarrow \chi_i f_\alpha} \rangle \Big) 
\bigg]
\,,\label{eq:BME12}
  \end{align}
  \begin{align}
   & x \mathcal{H}Z \,\frac{\dd Y_{\chi_i}}{\dd x} =
   \sum_\alpha \Bigg[ \langle\Gamma_{\phi\rightarrow \chi_i f_\alpha} \rangle Y_\phi + 
   \langle\Gamma_{\bar{\phi}\rightarrow \chi_i \bar{f}_\alpha} \rangle Y_{\bar{\phi}}
       \nonumber \\ &  \qquad 
  -Y_\phi^\text{eq}
\frac{Y_{\chi_i}}{Y_{\chi_i}^\text{eq}} \left( 
\frac{Y_{f_\alpha}}{Y_{f_\alpha}^\text{eq}}\, \langle\Gamma_{\bar{\phi}\rightarrow \chi_i \bar{f}_\alpha} \rangle  +\frac{Y_{\bar{f}_\alpha}}{Y_{f_\alpha}^\text{eq}}\,\langle\Gamma_{\phi\rightarrow \chi_i f_\alpha} \rangle \right) 
\Bigg] \,,
\label{eq:BME13} 
  \end{align}
where $\mathcal{H}$ is the Hubble expansion rate
and $\langle \Gamma_\phi \rangle=\langle\sum_{i,\alpha}\Gamma_{\phi\rightarrow \chi_i f_\alpha} \rangle$ is the total decay rate averaged over time dilation factors with the equilibrium phase-space distribution.

Note that for small asymmetries and $Y_{f_\alpha}^\eq\simeq Y_{f_\beta}^\eq$, we can approximate the right-hand side of Eq.~\eqref{eq:BME13}  [and the corresponding term in Eq.~\eqref{eq:BME11}] by 
$\left(\frac{Y_\phi + Y_{\bar{\phi}}}{2} -Y_\phi^\text{eq}\,
\frac{Y_{\chi_i}}{Y_{\chi_i}^\text{eq}}\right)
\sum_\alpha\Big(\langle\Gamma_{\phi\rightarrow \chi_i f_\alpha} \rangle  + 
   \langle\Gamma_{\bar{\phi}\rightarrow \chi_i \bar{f}_\alpha} \rangle \Big)$.
Furthermore, considering the limit of massless SM fermions,\footnote{More precisely, as this only concerns the Boltzmann equation for the asymmetry, which we only evolve down to $T_\text{sph}$, this limit practically implies $m_f\ll T_\text{sph}$, which is well fulfilled for all SM fermions but the top quark. } such that their equilibrium densities are independent of the SM flavor index $\alpha$, and
introducing 
\begin{align}
\label{eq:Gamma_i}
    \Gamma_i &\equiv \frac12\sum_\alpha\Big(\langle\Gamma_{\phi\rightarrow \chi_i f_\alpha} \rangle  + 
   \langle\Gamma_{\bar{\phi}\rightarrow \chi_i \bar{f}_\alpha} \rangle \Big),
   \\
    \Gamma_\phi & \equiv \sum_i \Gamma_i\,,
   \\
   \epsilon_i &\equiv\frac{1}{2 \Gamma_i }\,\sum_\alpha\Big(\langle\Gamma_{\phi\rightarrow \chi_i f_\alpha} \rangle -\langle\Gamma_{\bar{\phi}\rightarrow \chi_i \bar{f}_\alpha} \rangle   \Big),
\end{align}
as well as $\Sigma_\phi \equiv Y_\phi + Y_{\bar{\phi}}$, $\Delta_\phi\equiv Y_\phi - Y_{\bar{\phi}}$ and $\Delta_f\equiv\sum_\alpha (Y_{f_\alpha} - Y_{\bar{f}_\alpha})$, we can rewrite the Boltzmann equations in the compact form:
\begin{align}
  x \mathcal{H}Z \,\frac{\dd \Sigma_{\phi}}{\dd x} =&
  -\Gamma_\phi\,  \Sigma_{\phi}    
  +  
 \, \Sigma_\phi^\text{eq}  \, 
\sum_{i} \Gamma_i\,\frac{Y_{\chi_i}}{Y_{\chi_i}^\text{eq}}  
       \nonumber \\ & 
-\frac{1}{2}\langle\sigma v\rangle\, s\,\Big( \Sigma_{\phi}^2 - \Sigma_\phi^{\text{eq}2}\Big) \label{eq:BME21}
\\
    x \mathcal{H}Z \,\frac{\dd \Delta_{\phi}}{\dd x} 
   =&
  -\Gamma_\phi\,  \Delta_{\phi} 
     \nonumber \\ & 
+ \, \Sigma_\phi^\text{eq}  \, 
\sum_{i}\Gamma_i \, \frac{Y_{\chi_i}}{Y_{\chi_i}^\text{eq}} 
\bigg(\!-\epsilon_i+
\frac{\Delta_{f}}{\Sigma_{f}^\text{eq}}
\bigg)\label{eq:BME22}
\\
    x \mathcal{H}Z \,\frac{\dd Y_{\chi_i}}{\dd x} =&
\,\Gamma_i\left(\Sigma_\phi -\Sigma_\phi^\text{eq}\,
\frac{Y_{\chi_i}}{Y_{\chi_i}^\text{eq}}\right)\label{eq:BME23}
  \end{align}
where [in Eq.~\eqref{eq:BME22}] we can express $\Delta_f$ by $\Delta_\phi$ according to the respective relation among Eqs.~\eqref{eq:Deltae}--\eqref{eq:Deltad}.

Due to $CPT$ invariance the $\epsilon_i$ are not fully independent of each other. In particular, in the
minimal working example involving two dark flavors, $i=1,2$, we find $\epsilon_1 = - \epsilon_2 \,\mathcal{B}_2/\mathcal{B}_1\equiv\epsilon$ and we use $\bar\lambda$ (governing the overall size of $\Gamma_\phi$), 
$\epsilon$, and the branching ratio $\mathcal{B}_1 = 1- \mathcal{B}_2 =\Gamma_1/\Gamma_\phi$, to parametrize the input quantities the Boltzmann equations depend upon.

Note that by dividing Eq.~\eqref{eq:BME22} by $\epsilon_1$ we can obtain a solution for $\Delta_{\phi}/\epsilon$ for arbitrary $\epsilon$, thereby factorizing the dynamics of the particle densities from the computation of the $CP$ asymmetry.

 Since the symmetric contributions of conversion processes via scatterings off SM particles enter in the same way in the Boltzmann equations as the ones via (inverse) decays, we can include the former by simply replacing $\Gamma_i \to \Gamma_i + \Gamma_{\text{scat},\,i}$ in the respective terms, where 
{\begin{equation}
      \Gamma_{\text{scat},\,i} = \frac12\sum_{k,m}
    \Big(\langle\sigma_{\phi \overline{k} \to \chi_i m} v \rangle   + 
  \langle\sigma_{\overline{\phi} k \to \chi_i \overline{m}} v \rangle \Big)\,n_k^{\rm eq}.
\end{equation}

The scattering processes can be affected by soft and $t$-channel divergences that require regularization. Furthermore, thermal effects can induce kinematic blocking of decay channels, while scatterings can also contribute to the generation of the $CP$ asymmetry via
\begin{equation}
   \epsilon_{\text{scat},\,i} = \frac{1}{2 \Gamma_{\text{scat},\,i} }\,\sum_{k,m}\Big(\langle\sigma_{\phi \overline{k} \to \chi_i m} v \rangle   -
  \langle\sigma_{\overline{\phi} k \to \chi_i \overline{m}} v \rangle \Big)\,n_k^{\rm eq}.
\end{equation}
A complete finite-temperature treatment of all these effects is beyond the scope of this work.\footnote{See Ref.~\cite{Becker:2023vwd} for a recent account on thermal corrections to conversion processes in similar models.} Instead, we consider two benchmark setups that provide reasonable estimates and give an indication of the associated theoretical uncertainty:
\begin{itemize}
\item \textit{Minimal setup}:
We regulate the soft divergences of the scattering processes minimally by only including a thermal gluon mass and a thermal quark mass for light quarks in the phase-space integration. Scatterings are included in Eqs.~\eqref{eq:BME21} and~\eqref{eq:BME23}. In this setup, they contribute to maintaining thermal equilibrium between the DM and mediator sectors, but do not contribute to the generation of the $CP$ asymmetry.

\item \textit{Thermal-mass setup}:
We include thermal masses for all particles entering the scattering and decay processes as detailed in appendix~\ref{app:thermalcorrections}. We approximately account for the asymmetric part of the scattering processes by assuming $\epsilon_{\text{scat},\,i}\simeq\epsilon_i$ (see also ref. \cite{Pilaftsis:2003gt}), and include scatterings in all Boltzmann equations, Eqs.~\eqref{eq:BME21}--\eqref{eq:BME23}. This setup captures kinematic blocking effects, which can temporarily close the decay channel during the thermal evolution, while retaining scattering contributions in all evolution equations.
\end{itemize}

In the following, we focus on a minimal working realization with two DM flavors $\chi_{1,2}$ coupled to two flavors of right-handed down-type quarks, such that $\lambda$ is a $2\times 2$ matrix. The presence of two DM and two SM flavors is required for a non-vanishing $CP$ asymmetry, while this minimal setup is sufficient to capture the relevant dynamics. We choose down-type quarks such that the massless-fermion approximation employed in Eq.~\eqref{eq:BME22} is well justified. Within this approximation, the cosmological evolution does not depend on the specific choice of the two down-type flavors. Furthermore, as the mass splitting within the DM multiplet is expected to be extremely small (cf. Secs.~\ref{sec:model} and \ref{sec:viableparam}), for practical purposes, we set $m_{\chi_1}= m_{\chi_2}\equiv m_\chi$ in the numerical solution of the Boltzmann equations.

We compute the leading order cross sections for the conversion and mediator annihilation processes analytically using \textsc{FeynCalc}~\cite{Shtabovenko:2023idz, Shtabovenko:2020gxv, Shtabovenko:2016sxi, Mertig:1990an} and \textsc{FeynArts}~\cite{Hahn:2000kx}. To account for effects of (excited) bound states of mediator particles during freeze-out, we employ the package \mbox{\textsc{BSFfast}~\cite{Binder:2025daq}} based on Refs.~\cite{Garny:2021qsr,Binder:2023ckj} and we include Sommerfeld enhancement.

\begin{figure*}[t]  
    \centering
    \begin{minipage}{0.92\textwidth}
    \begin{subfigure}[c]{0.49\textwidth}
        \centering
        \qquad\quad Minimal setup\vspace*{.3mm}
        \includegraphics[width=\textwidth]{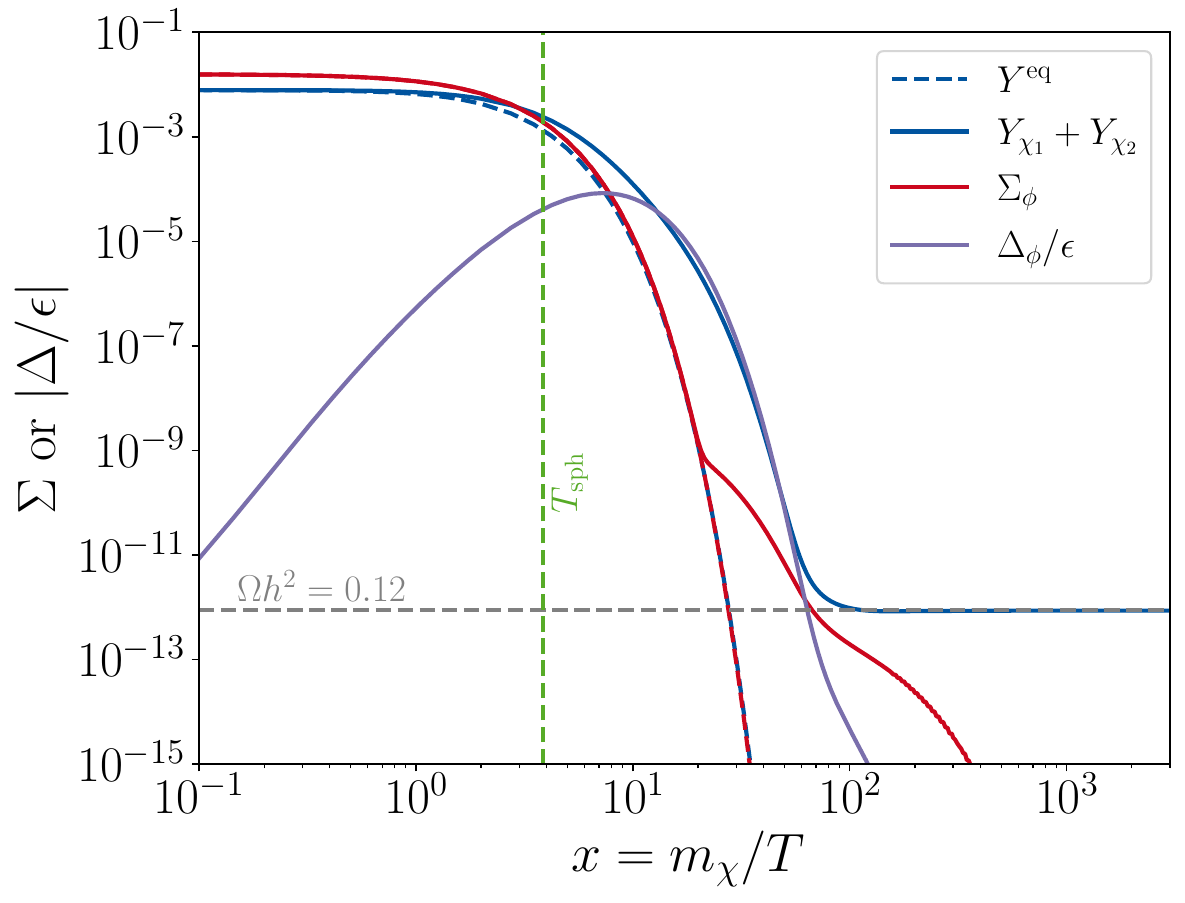}
    \end{subfigure}
    \begin{subfigure}[c]{0.49\textwidth}
        \centering
        \qquad\quad  Thermal-mass setup\vspace*{.3mm}
        \includegraphics[width=\textwidth]{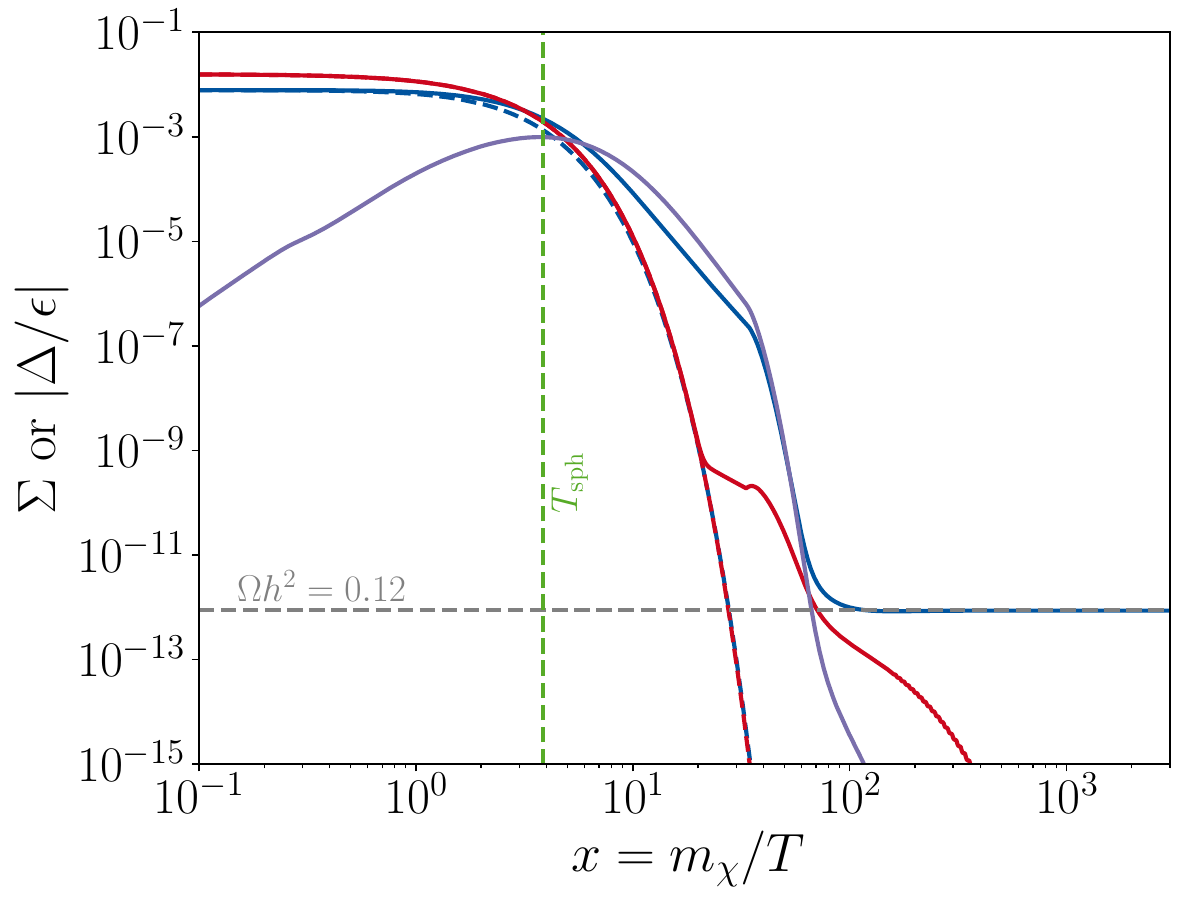}
    \end{subfigure}
    \end{minipage}
  \caption{Solutions of the Boltzmann equations for the parameter point $m_{\chi}=500$ GeV and $m_{\phi}=510$ GeV and $\mathcal{B}_1=0.75$. The left column corresponds to the minimal setup, while the right column corresponds to the thermal-mass setup. In each case, $\bar\lambda$ is chosen such that $\Omega h^2 = 0.12$. 
  }
  \label{fig:BME_solution_examples}
\end{figure*}

In Fig.~\ref{fig:BME_solution_examples} we show the thermal evolution of the particle abundances and asymmetry solving the Boltzmann equations for a DM mass of $m_{\chi}=500\,$GeV, 
and a mass splitting between $m_{\chi}$ and the mediator of $\Delta m = 10$  GeV\@. The branching ratio of the mediator decaying into $\chi_1$ is set to $\mathcal{B}_1=0.75$, and the left and right panels show the corresponding results employing the minimal and thermal-mass setup, respectively. In all cases, we set $\lambda_H=0$ and adjust the overall Yukawa coupling $\bar{\lambda}$ such that the model explains the measured relic abundance $\Omega h^2 = 0.12$ \cite{Planck:2018vyg}.  
The $y$-axis shows the comoving number densities of $\chi_1 + \chi_2$ (blue) and $\phi$ (red), as well as the asymmetry $\Delta_\phi/\epsilon$ (purple).
The corresponding equilibrium densities are shown as dashed lines. The green dashed vertical line indicates the sphaleron decoupling temperature $T_\text{Sph}$. 

In both the minimal and the thermal-mass setup, the asymmetry $\Delta_\phi/\epsilon$ starts to build up due to conversion processes of $\chi_1$ and $\chi_2$ into $\phi$ while the former fall out of thermal equilibrium. However, in the thermal-mass setup, the asymmetry starts to build up at earlier times (higher temperatures) compared to the minimal setup. This is because in the thermal-mass setup we take into account the scattering processes in the source term of the asymmetry, which increases the production of the asymmetry at earlier times. The thermal blocking effects in the thermal-mass setup lead to kinks in the evolution of the asymmetry, which are not present in the minimal setup. These kinks correspond to the points in time where the decay suddenly becomes kinematically forbidden or allowed again. 

\section{Viable parameter space}
\label{sec:viableparam}

\subsection{Relic density and baryon asymmetry}

We now determine the viable parameter space of the model in which both the observed DM relic abundance and the observed baryon asymmetry can be explained. To this end, we first compute the conversion-driven freeze-out boundary which gives us the maximum mass splitting $\Delta m$ for a given DM doublet mass $m_{\chi}$ in which the DM can still be produced via conversion-driven freeze-out. The region extends up to $m_{\chi} \sim 1300$ GeV and allows for mass splittings up to $\Delta m \sim 20\,$GeV. Then, for a grid of points in the $m_{\chi}$-$\Delta m$ plane below the conversion-driven freeze-out boundary and for a given branching ratio $\mathcal{B}_1$, we solve the Boltzmann equations while varying the overall Yukawa coupling $\bar{\lambda}$  to find the value of $\bar\lambda$ that explains the measured relic density.
The values of $\bar\lambda$ required to obtain the correct relic abundance are shown in the upper panels of Fig.~\ref{fig:lambda_values} for $\mathcal{B}_1=0.75$, $\lambda_H=0$, and, again, for both the minimal (left) and thermal-mass (right) setup. The white region above the color-coded one corresponds to the region where conversion-driven freeze-out is not possible and significantly larger couplings are required (WIMP region within the model). In all scenarios, we find that the required values of $\bar\lambda$ are in the range $10^{-7}\! -\! 10^{-6}$ and do not differ much when going from the minimal setup to the thermal-mass setup. 

\begin{figure*}[tbp]  
    \centering
    \begin{subfigure}[c]{0.49\textwidth}
        \centering
        \qquad\quad Minimal setup\vspace*{.3mm}
        \includegraphics[width=\textwidth]{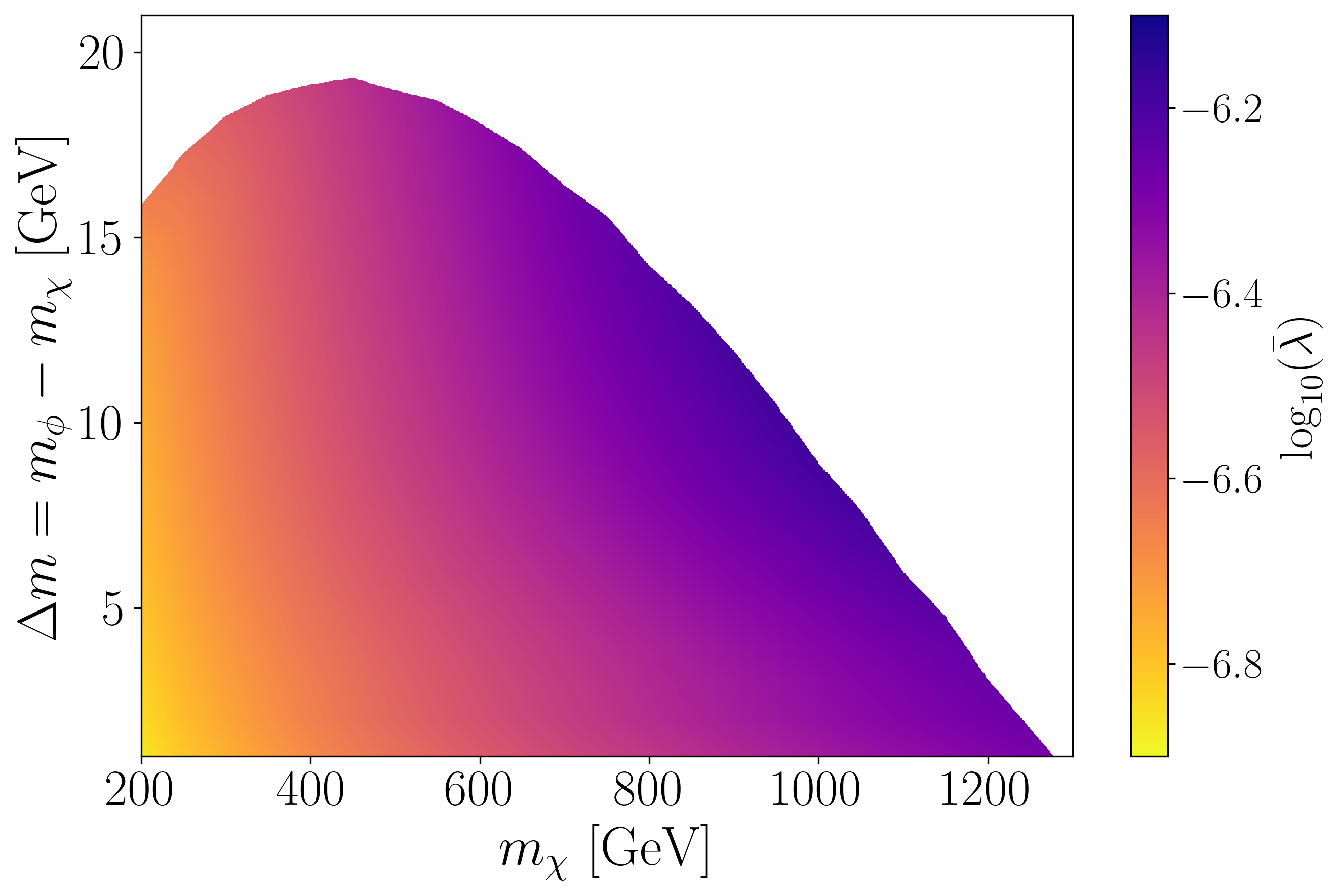}
    \end{subfigure}
    \begin{subfigure}[c]{0.49\textwidth}
        \centering
        \qquad\quad Thermal-mass setup\vspace*{.3mm}
        \includegraphics[width=\textwidth]{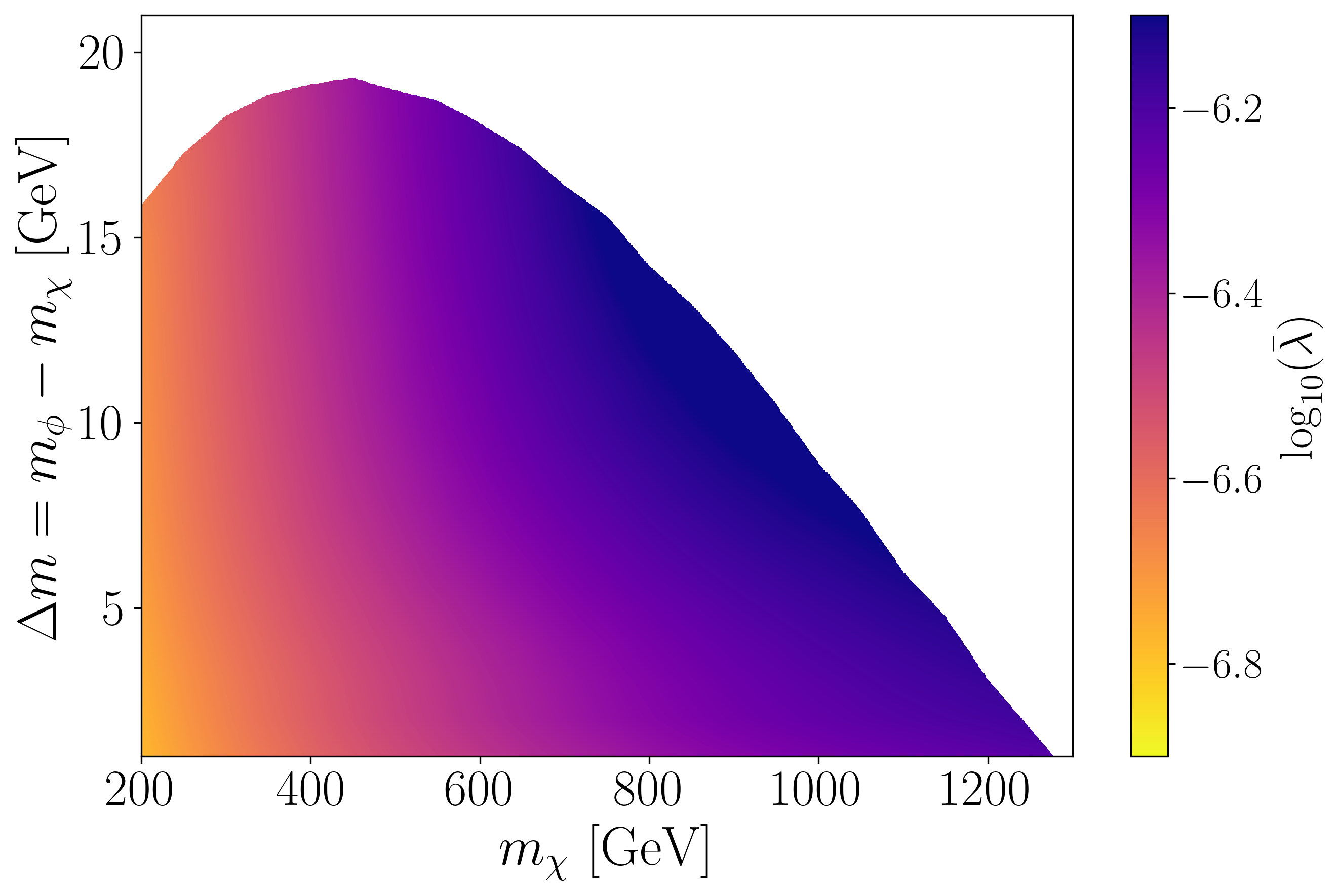}
    \end{subfigure}
    \\
    \begin{subfigure}[c]{0.49\textwidth}
        \centering
        \includegraphics[width=\textwidth]{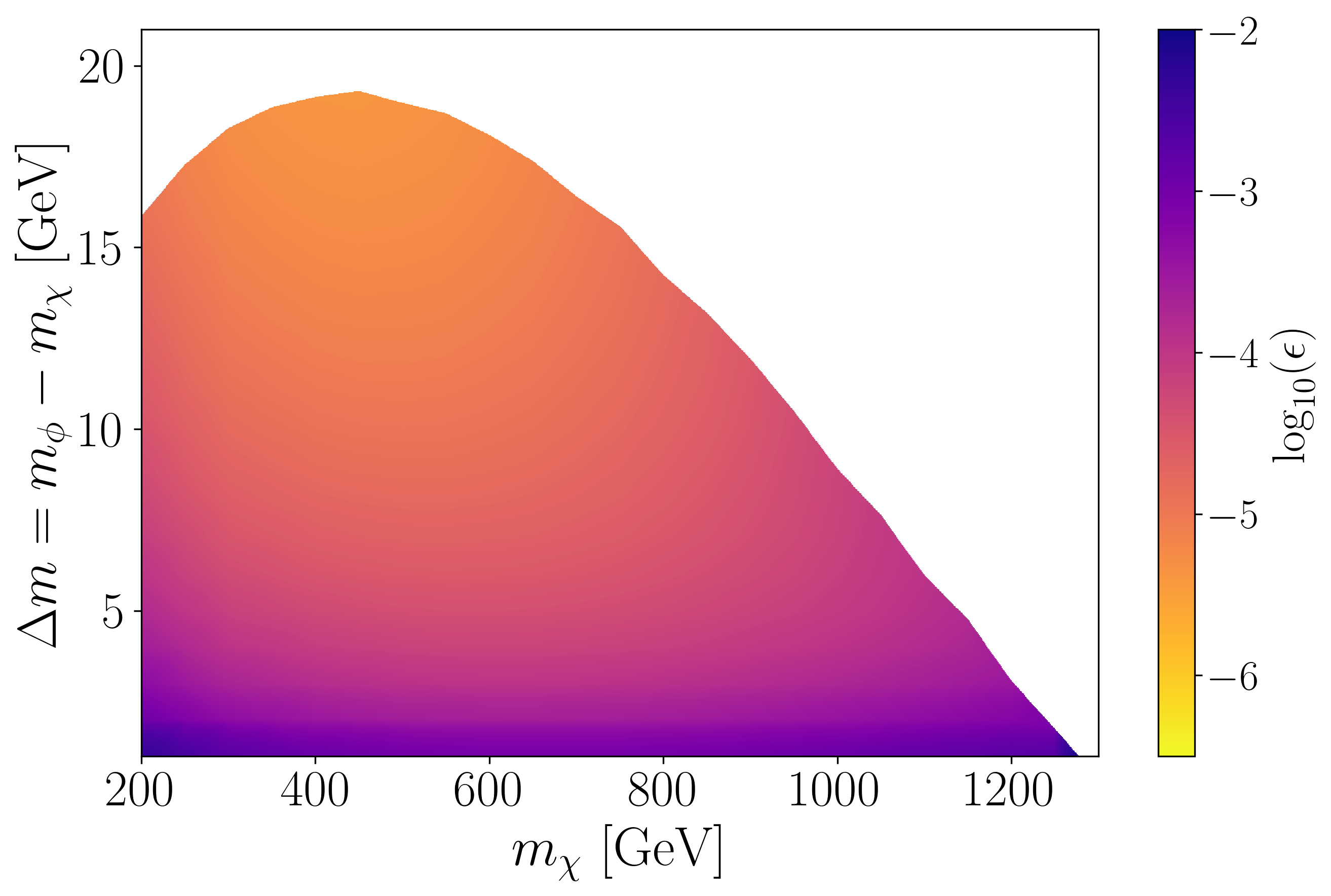}
    \end{subfigure}
    \begin{subfigure}[c]{0.49\textwidth}
        \centering
        \includegraphics[width=\textwidth]{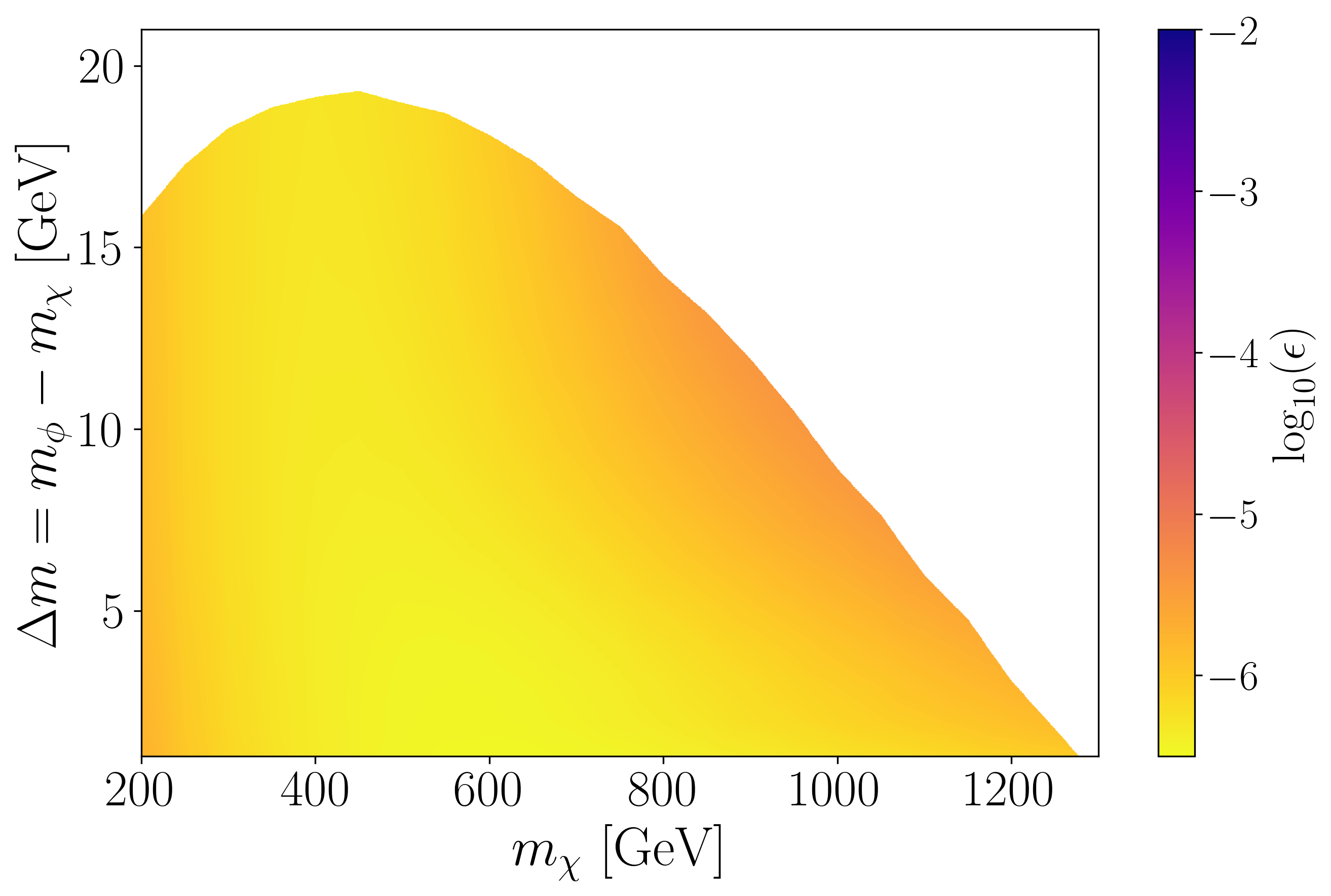}
    \end{subfigure}
    \\
    \begin{subfigure}[c]{0.49\textwidth}
        \centering
        \includegraphics[width=\textwidth]{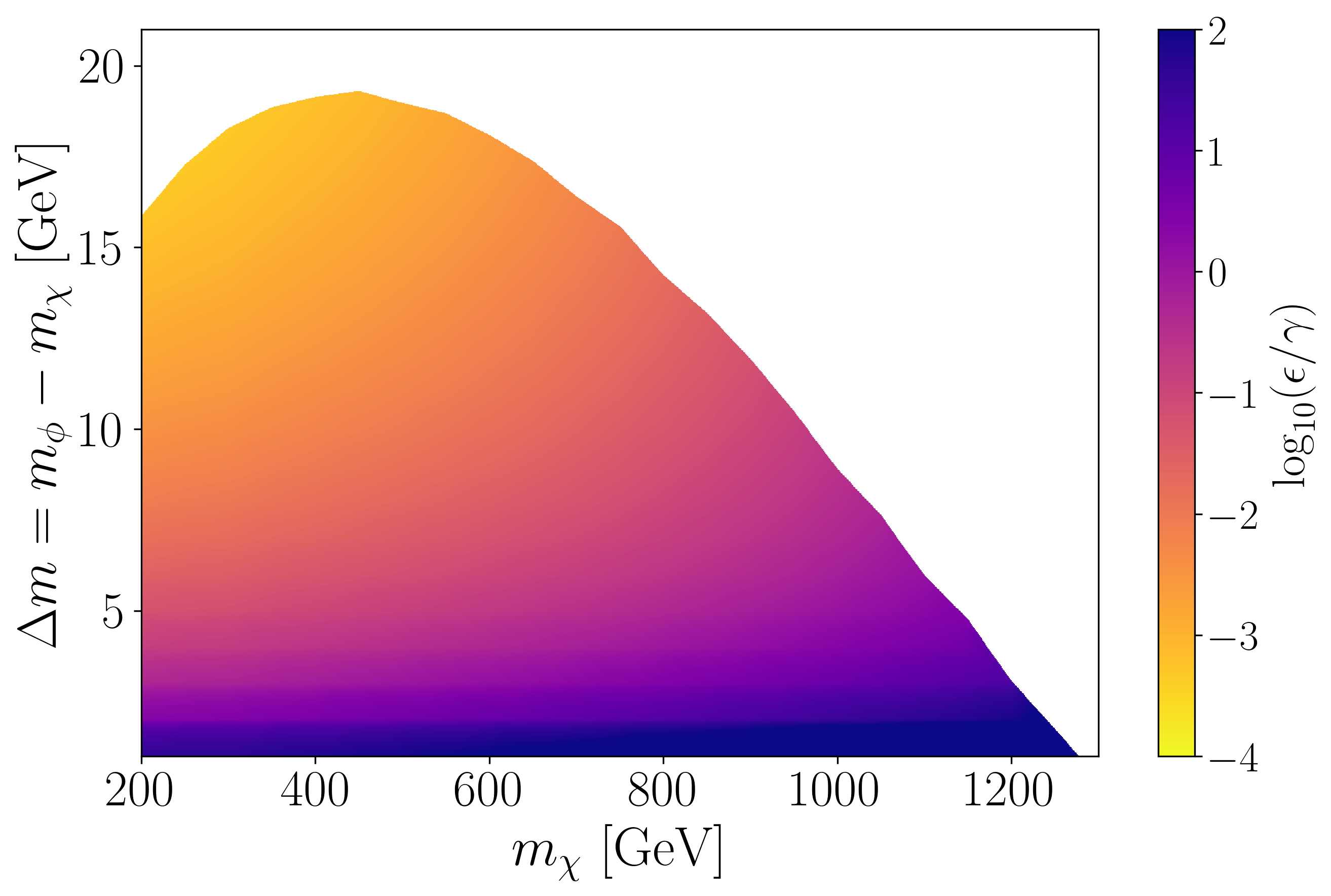}
    \end{subfigure}
    \begin{subfigure}[c]{0.49\textwidth}
        \centering
        \includegraphics[width=\textwidth]{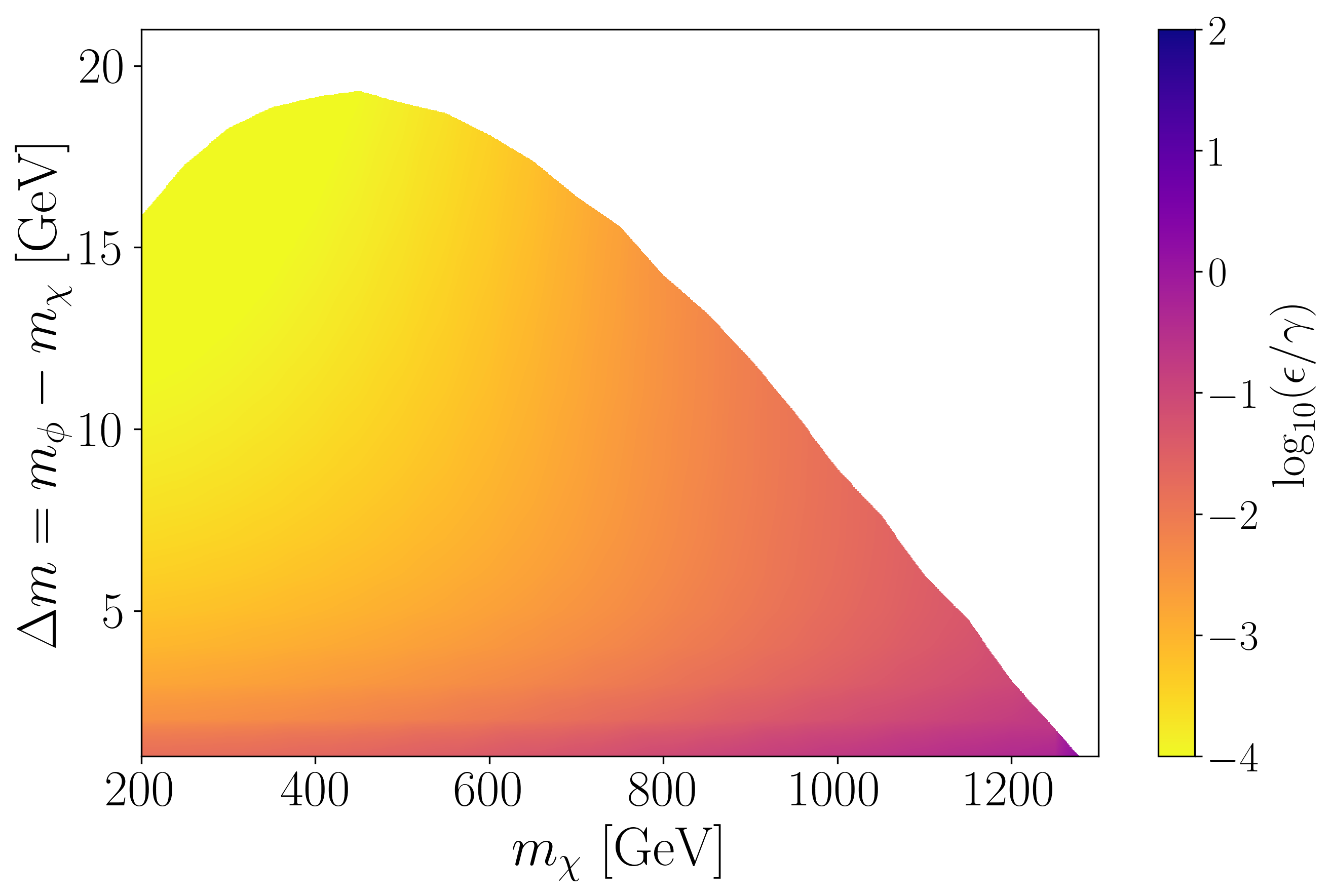}
    \end{subfigure}
  \caption{Results for the Yukawa coupling 
    $\bar\lambda$ (upper panels), the $CP$ asymmetry $\epsilon$ (middle panels), and the ratio $\epsilon/\gamma$ (lower panels) required to obtain the correct relic abundance and BAU in the conversion-driven freeze-out region for $\mathcal{B}_1=0.75$, $\lambda_H=0$ and for the minimal (left panels) and thermal-mass setup (right panels).}
    \label{fig:lambda_values}
\end{figure*}

Then, for each point in the $m_{\chi_1}$-$\Delta m$ plane, we infer the $CP$ asymmetry $\epsilon$ needed to obtain the correct baryon asymmetry. This is done by employing the solutions of the Boltzmann equations for $\Delta_\phi/\epsilon$ at the sphaleron decoupling temperature $T_\text{Sph}$ and using the relation of Eq.~\eqref{eq:nb_down} in the symmetric phase to compute the baryon asymmetry $Y_{\Delta B}$ and compare it to the observed value $Y_{\Delta B}^\text{meas} \simeq 0.9\times 10^{-10}$~\cite{Davidson:2008bu,Planck:2018vyg}. 
We find viable solutions that reproduce $Y_{\Delta B}$ for values of $\epsilon$ that range from $\sim 10^{-6}$ to $\sim 10^{-3}$ in the minimal setup and from $\sim 10^{-7}$ to $\sim 10^{-6}$ in the thermal-mass setup, see the middle panels of Fig.~\ref{fig:lambda_values}.

\begin{figure}[tb]
    \centering
    \includegraphics[width=0.6\linewidth]{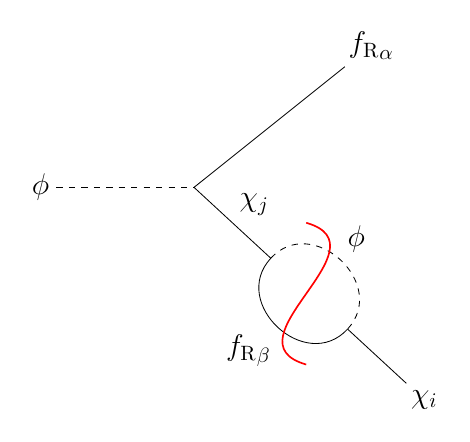}
    \caption{One-loop self-energy diagram with thermal cut leading to the $CP$-asymmetry in the mediator decay $\phi \to \chi_i {f_{\mathrm{R}}}_\alpha$.}
    \label{fig:phi-decay}
\end{figure}

At first sight, the required values of $\epsilon$ appear large given the small Yukawa couplings $\bar\lambda\sim 10^{-7}$--$10^{-6}$ that characterize the conversion-driven freeze-out regime and enter the $CP$-violating conversion processes. However, achieving such values is made possible by a resonant enhancement of the $CP$ asymmetry when the two DM states are nearly degenerate, as naturally expected and considered in our scenario. We therefore allow for a small splitting $\Delta m_{12}=m_{\chi_2}-m_{\chi_1}$ between the two DM flavors. When this splitting is sufficiently small -- naturally in the ballpark of $\bar\lambda^2m_\chi$ -- the self-energy contribution to the conversion process is enhanced, allowing for a sizable asymmetry despite the small overall Yukawa coupling.

The main source of ${CP}$ violation in this model are thermal corrections to (inverse) $\phi$~decays.
While $CP$ is conserved at zero temperature, the interference of the tree-level $\phi \to {f_{\mathrm{R}}}_\alpha \chi_i$ diagram and the corresponding one-loop diagram with a quark self-energy correction, shown in Fig.~\ref{fig:phi-decay},
generates an absorptive part due to thermal corrections, in which case the~$\phi$ or~${f_{\mathrm{R}}}_\beta$ running in the loop can originate from the thermal bath.
The amount of ${CP}$ violation generated is parameterized by~$\epsilon(T)$ and has been computed in Ref.~\cite{Hambye:2016sby} for the case of the SM Higgs doublet decaying into a lepton and a sterile neutrino, which we can adopt here as well, accounting for the correct rescaling due to the different $\mathrm{SU}(N)$ groups (see below). 
Following the notation of Ref.~\cite{Heisig:2024mwr} it reads
\begin{align}
    \epsilon(T)
    =
    I_1  \frac{\xi \gamma(T)}{\left[\xi+\pi\varphi/(2x)^2\right]^2 + {\gamma(T)}^2} 
    \simeq
    I_1  \frac{\gamma(T)}{\xi}
    \,,
    \label{eq:epsilon}
\end{align}
where we have
\begin{align}
    I_1
    &= \frac{\mathrm{Im}\!\left[(\lambda^\dagger \lambda)_{12}^2\right]}{(\lambda^\dagger \lambda)_{11} (\lambda^\dagger \lambda)_{22}}
    =\frac{\mathrm{Im}\!\left[(\lambda^\dagger \lambda)_{12}^2\right]}{{\bar{\lambda}}^4 \mathcal{B}_1 (1-\mathcal{B}_1) }
    \,,
    &
    \xi
    &=
    2\frac{\Delta m_{12}}{\Gamma_{\chi_2}^0}
    \,,
    \label{eq:I1}
\end{align}
and $\varphi$ is an $\mathcal{O}(1)$~number parameterizing thermal corrections to the DM mass splitting~$\Delta m_{12}$.
The quantity $\Gamma_{\chi_2}^0$ denotes the would-be decay width of $\chi_2\to\phi f$ in the limit of vanishing final-state masses, whereas $\gamma(T)$ denotes the spectral loop integral entering the self-energy diagram (see appendix~D of Ref.~\cite{Frossard:2012pc} for details), such that their product $\Gamma_{\chi_2}^0 \gamma(T)$ corresponds to the absorptive part of the self-energy.
The only difference with respect to Refs.~\cite{Frossard:2012pc,Hambye:2016sby,Heisig:2024mwr} is a relative factor of~$3/2$ for~$\gamma$ to account for the correct $\mathrm{SU}(N)$ factors.
The approximation on the right-hand side of Eq.~\eqref{eq:epsilon} is obtained in the limit $\xi \gg \gamma(T),\, x^{-2}$, which holds in the semi-relativistic regime $x \gtrsim 2$ and assuming $\Delta m_{12} \gg \Gamma_{\chi_2}^0$ as well as using that $\gamma(T)$ is a steeply falling function in $x=m_{\chi_1}/T$.

Dividing Eq.~\eqref{eq:epsilon} by $\gamma$ yields 
\begin{equation}
\label{eq:egIxi}
    \frac{\epsilon}{\gamma} \simeq  \frac{I_1}{\xi}
    \,,
\end{equation} 
where the right-hand side is a temperature-independent function of the model parameters, particularly $\Delta m_{12}/m_{\chi_2}$ and the complex phases of $\lambda$. 
This allows us to gain insights into the viable parameter space explaining the BAU without explicitly sampling actual realizations of $\lambda, \Delta m_{12}$ and computing $\epsilon(T)$ via Eq.~\eqref{eq:epsilon}. Instead, for each point in the so-far analyzed parameter space characterized by $m_{\chi_1}, \Delta m, \mathcal{B}_1$ and corresponding value of $\bar\lambda$ that obeys $\Omega h^2 = 0.12$, we compute $\gamma(T_\text{Sphal})$ and with the respective $\epsilon(T_\text{Sphal})$, we compute $\epsilon/\gamma$. This quantity is shown in the lower panels of Fig.~\ref{fig:lambda_values}, and thanks to the approximate relation, Eq.~\eqref{eq:egIxi}, admits a more direct interpretation than $\epsilon$ itself: while the latter depends on the size of the thermal spectral function, $\epsilon/\gamma$ directly probes the combination $ I_1/\xi$ of temperature-independent model parameters that controls whether the required asymmetry can be realized.

A realistic upper limit of this quantity's range can be assessed as follows: On the one hand, using the definition of $I_1$ in Eq.~\eqref{eq:I1} and the Cauchy-Schwarz inequality shows $|I_1| \leq 1$. So far our analysis only constrained the matrix $\lambda$ via its two degrees of freedom $\bar\lambda$ and $\mathcal{B}_1$ that entered the Boltzmann equations,
theoretically 
leaving enough freedom to saturate the above limit.\footnote{ Particularly, any rank-one matrix $\lambda$ with $\arg[(\lambda^\dagger \lambda)_{12}]=\pm \pi/4$ saturates this limit as can be seen considering the parametrization
\begin{align}
    \lambda^\dagger \lambda
    &=
    {\bar{\lambda}}^2 \begin{pmatrix}
        \mathcal{B}_1 & e^{i\theta} \sqrt{\mathcal{B}_1 (1\!-\!\mathcal{B}_1) \!-\! D}
        \\
        e^{-i\theta} \sqrt{\mathcal{B}_1 (1\!-\!\mathcal{B}_1)\! -\! D} & (1-\mathcal{B}_1)
    \end{pmatrix}\,,
\end{align}
where $D \equiv \det(\lambda^\dagger \lambda)/{\bar\lambda}^4$ and $\theta$ is the complex phase.}

On the other hand, the above computation assumes $\xi \gg 1$. We therefore focus on the region $\epsilon/\gamma < 0.1$, which corresponds to values of $ I_1/\xi$ that can be achieved without pushing this hierarchy too strongly. 
This should not be understood as a sharp physical exclusion criterion: viable solutions may well exist for larger values of $\epsilon/\gamma$. However, in that regime the two nearly degenerate DM states may no longer be adequately described as incoherent particle species, and a more complete treatment including mixing and density-matrix effects would be required to establish the viability quantitatively.

The viable parameter space is shown in Fig.~\ref{fig:parameter_space} for the two different thermal treatments and the two different branching ratios $\mathcal{B}_1$ in terms of the DM mass $m_{\chi_1}$ and the mass splitting $\Delta m$. In black, we show the conversion-driven freeze-out boundary under which we find viable solutions for the relic abundance and the baryon asymmetry in the entire region. The parameter space that corresponds to $\epsilon/\gamma > 0.1$ is shown in gray. This region covers a significant portion of the parameter space in the minimal setup, while in the thermal-mass setup, the viable parameter space is mostly within the region where $\epsilon/\gamma < 0.1$ because the values of $\epsilon$ needed to obtain the correct baryon asymmetry are smaller as the asymmetry starts to build up at earlier times. Because of the very small Yukawa couplings needed to obtain the correct relic abundance, the mediator $\phi$ is long-lived with decay lengths that can reach up to $\mathcal{O}(1)$ m. The lifetime of the mediator is shown as different contours in blue in the $m_{\chi_1}$-$\Delta m$ plane. 

So far, we have chosen $\lambda_H=0$, corresponding to a vanishing Higgs-portal interaction of the mediator. A non-zero $\lambda_H$ opens additional mediator-pair annihilation channels and can therefore, in principle, enlarge the viable parameter space. In the quark-philic scenario considered here, however, mediator annihilation is typically dominated by QCD interactions including bound-state effects. We have checked that even for a sizable Higgs-portal coupling, $\lambda_H=1$, the impact is mild: at large $\Delta m$, the CDFO boundary is shifted slightly towards larger mass splittings, while towards small mass splittings the parameter space is virtually unchanged. The qualitative picture therefore remains the same. This is in contrast to the leptophilic case, where the Higgs-portal interaction can play a much more important role~\cite{Heisig:2024mwr}.

\subsection{LHC signatures and constraints}
\label{sec:LHC}

\begin{figure*}[tbp]  
    \centering
  \begin{tabular}{@{}c@{\hspace{0.5em}}c@{}}
    \begin{tabular}{@{}c@{}}
      \adjustbox{angle=90,valign=c}{$\mathcal{B}_1=0.75$} \\
      [0.2\textwidth]
      \adjustbox{angle=90,valign=c}{$\mathcal{B}_1=0.95$}
    \end{tabular}
    &
    \begin{minipage}{0.92\textwidth}
    \begin{subfigure}[c]{0.49\textwidth}
        \centering
        \qquad\quad Minimal setup\vspace*{.3mm}
        \includegraphics[width=\textwidth]{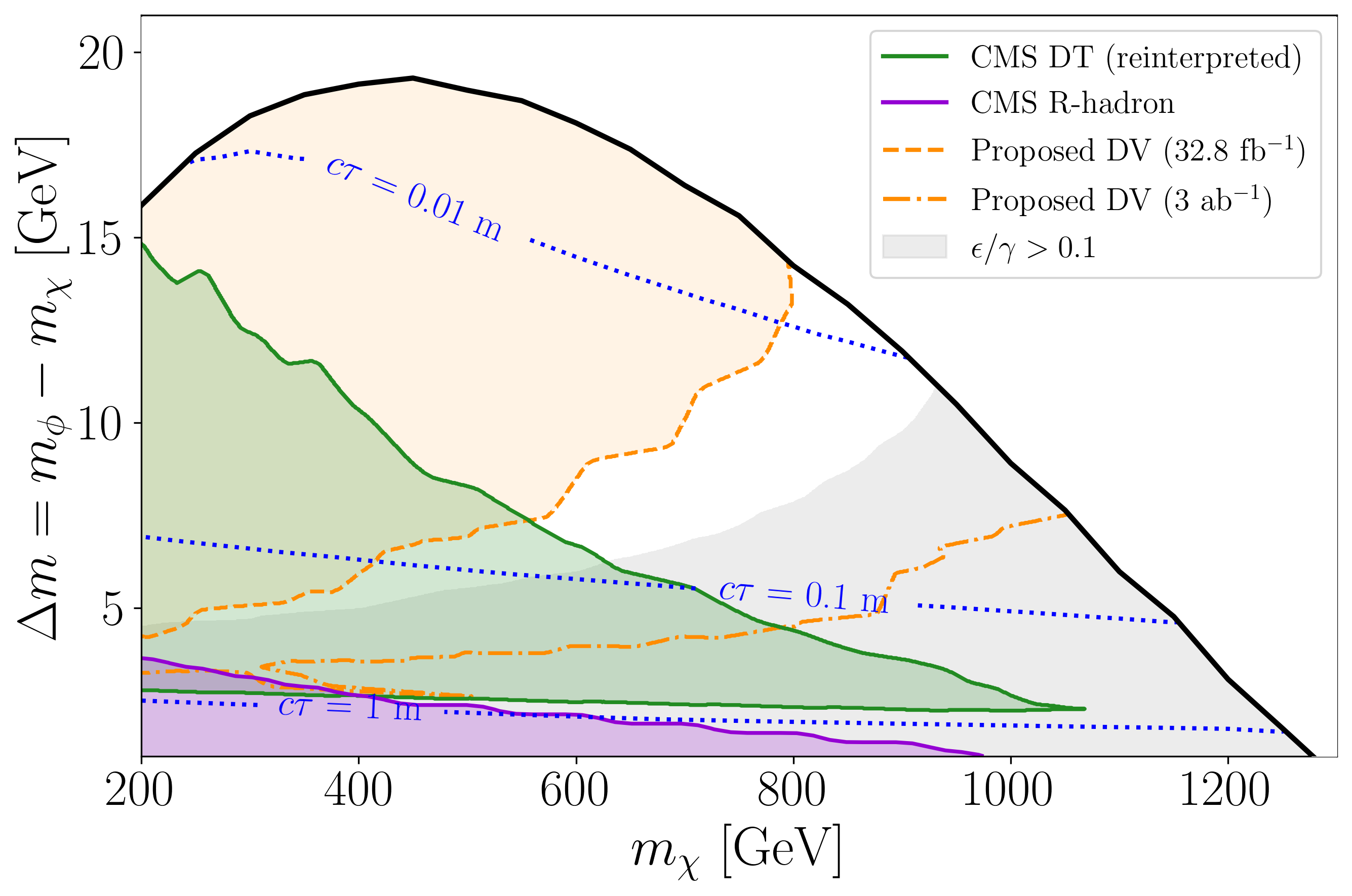}
    \end{subfigure}
    \begin{subfigure}[c]{0.49\textwidth}
        \centering
        \qquad\quad Thermal-mass setup\vspace*{.3mm}
        \includegraphics[width=\textwidth]{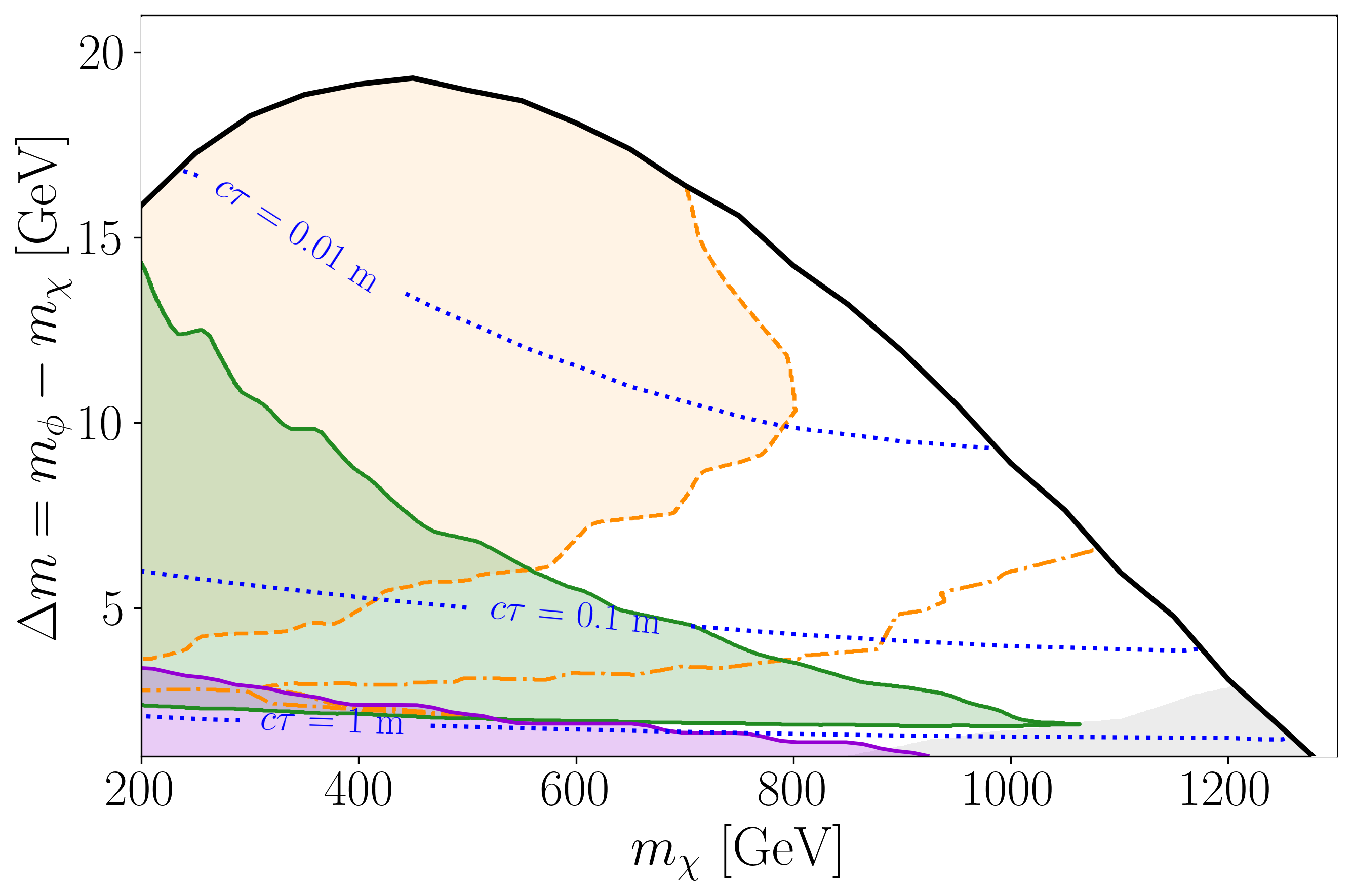}
    \end{subfigure}
    \\
    \begin{subfigure}[c]{0.49\textwidth}
        \centering
        \includegraphics[width=\textwidth]{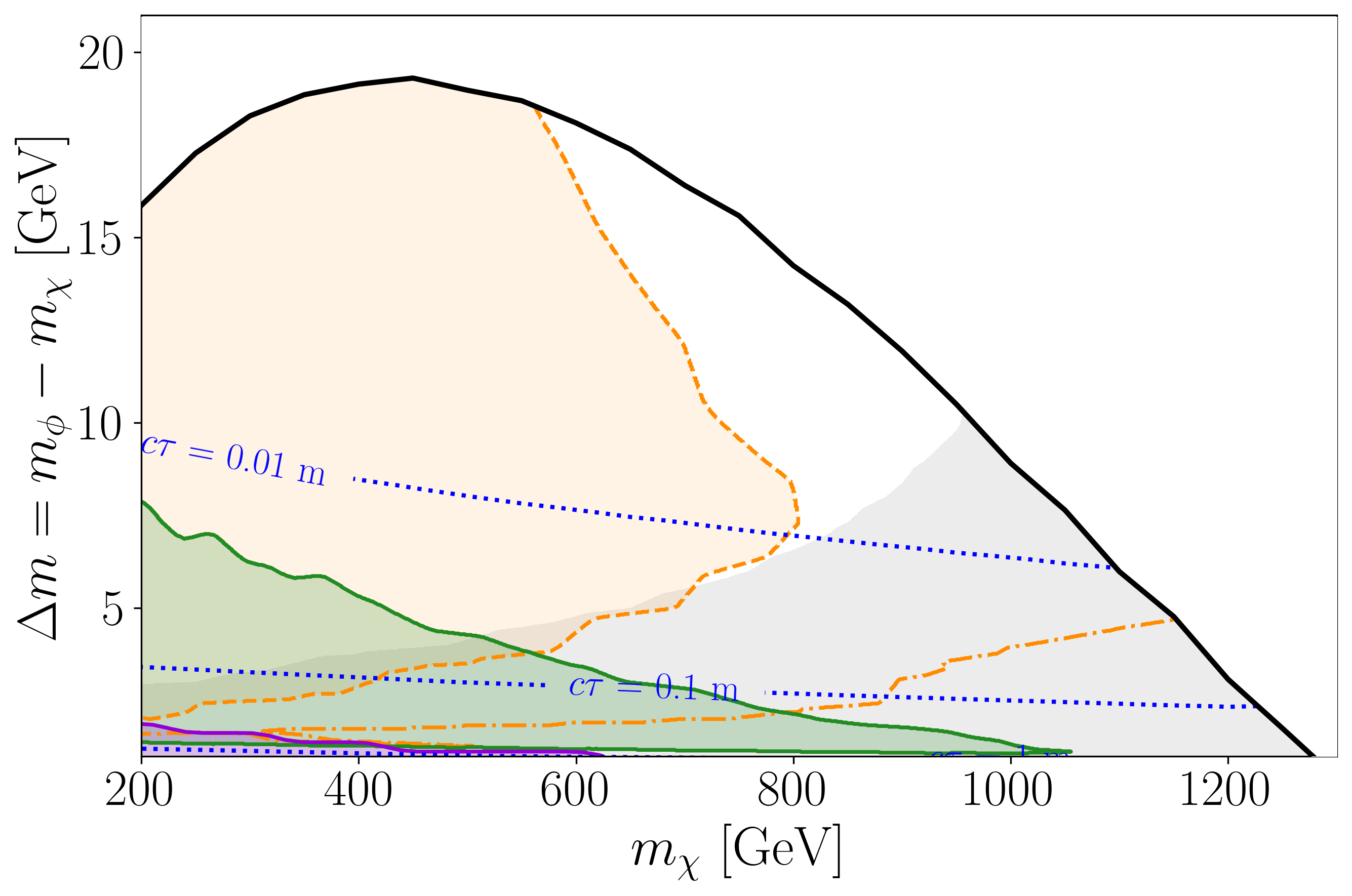}
    \end{subfigure}
    \begin{subfigure}[c]{0.49\textwidth}
        \centering
        \includegraphics[width=\textwidth]{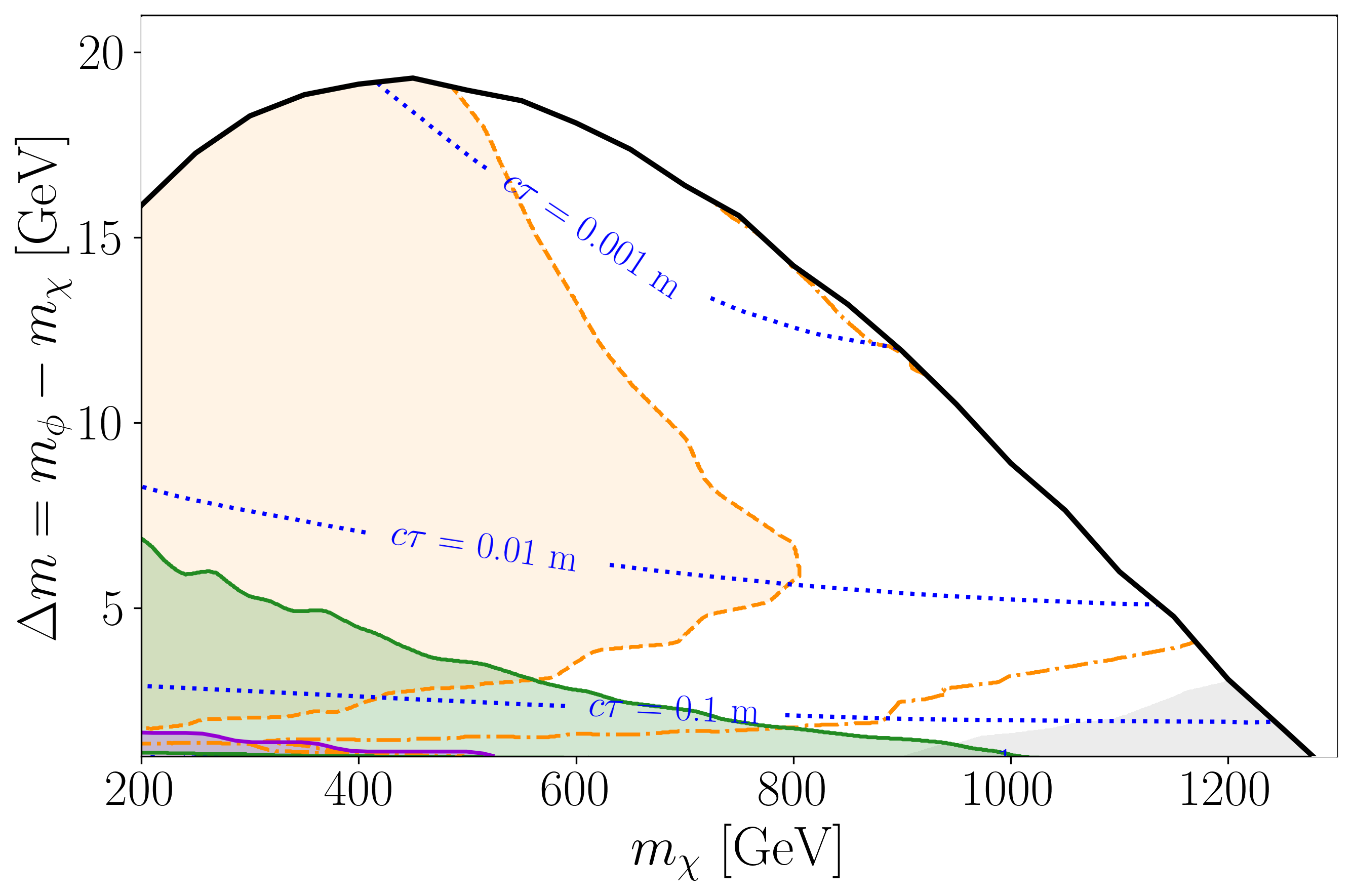}
    \end{subfigure}
    \end{minipage}
  \end{tabular}
  \caption{Viable parameter space below the CDFO boundary that provides correct relic abundance and baryon asymmetry. The green and purple shaded regions show the current limits from LHC searches for long-lived particles. The light orange shaded region displays the reach  of a proposed displaced-vertex search. The gray region corresponds to the region where $\epsilon/\gamma > 0.1$ and the blue contours show the lifetime of the mediator $\phi$.}
  \label{fig:parameter_space}
\end{figure*}

The long-lived mediator $\phi$ can be probed at the LHC via searches for disappearing tracks, displaced vertices, and heavy stable charged particles (HSCPs). Larger decay lengths of $\mathcal{O}(1)$ m can be probed via HSCP searches where the long-lived mediator $\phi$ would traverse the entire detector and be identified using time-of-flight and energy loss measurements. For intermediate decay lengths of $\mathcal{O}(10)$ cm to $\mathcal{O}(1)$ m, the mediator $\phi$ can be probed via disappearing track searches where the long-lived particle leaves a track in the inner detector that suddenly disappears due to the decay of the particle. For shorter decay lengths smaller than $\mathcal{O}(10)$ cm, displaced vertex searches can be used by looking for a vertex that is displaced from the primary interaction point. Because of the very small Yukawa couplings needed, constraints from the LHC are the most relevant for probing the viable parameter space as canonical direct and indirect detection searches are not sensitive to this scenario. 

Flavor and EDM observables likewise do not impose relevant constraints on the model, owing to the small couplings $\bar{\lambda}\lesssim 10^{-6}$.
The most sensitive flavor observable in the down-type quark sector is $\epsilon_K$ parameterizing the $CP$~violation in \smash{$K^0-\overline{K}^0$} mixing.
The constraints on the coupling matrix~$\lambda$ imposed by the current measurements of~$\epsilon_K$~\cite{ParticleDataGroup:2018ovx} in the given model have been derived in Eq.~(B.56) of Ref.~\cite{Belfatto:2025ids}.
Using this result, we find that the present model is more than 16~orders of magnitude below the current sensitivity reach of~$\epsilon_K$.
Also electric dipole moments (EDMs) are strongly suppressed in this model due to the chiral structure of the dark sector, coupling exclusively to right-handed down-type quarks at tree level. 
This forbids one-loop contributions to EDMs in the model~\cite{Agrawal:2014aoa} in addition to the very small coupling.
Thus, neither flavor nor EDM measurements provide relevant additional constraints in the parameter region of interest.

We now turn to the existing collider constraints, shown as the green and purple shaded regions in
Fig.~\ref{fig:parameter_space}. 
The green one denotes the 95\% CL exclusion  from the CMS searches for disappearing tracks (DT)~\cite{CMS:2018rea,CMS:2020atg}. To derive it, we reinterpret the analysis of Ref.~\cite{Heisig:2024xbh} performed in a similar scenario by using the obtained limits as function of the mediator mass and lifetime.  
The purple shaded area denotes the exclusion from HSCP searches  which we derived by using SModelS~\cite{Alguero:2021dig}. For the relevant parameter points, the strongest constraint comes from the CMS HSCP search for detector-stable $R$-hadrons in Ref.~\cite{CMS:2016ybj}. For a given mediator mass and lifetime, SModelS approximately reinterprets the corresponding search result by estimating the fraction of events containing at least one mediator particle that traverses the relevant parts of the detector before decaying~\cite{Heisig:2015yla,Heisig:2018kfq}. The disappearing-track search provides the strongest existing constraint over a larger region at low and intermediate DM masses, but loses sensitivity for $m_\chi$ approaching $1\,\mathrm{TeV}$ while HSCP searches exclude only a small region at very small mass splittings.

Other searches, such as missing energy and existing disappearing track searches (not shown here) are of limited reach and do not exceed $m_\chi\sim 400\,$GeV for such small mass splittings $\Delta m$~\cite{Heisig:2024xbh}  leaving ample allowed parameter space, particularly in the upper right half of the conversion-driven freeze-out region. 

This region can, however, be probed at the (HL-)LHC with dedicated strategies. To this end we display the projected reach of a displaced vertices (DV) search proposed in Ref.~\cite{Heisig:2024xbh} dedicated to probe the soft displaced signatures of conversion-driven freeze-out. The search is identical to the ATLAS DV search \cite{ATLAS:2017tny} but with a lower cut on the invariant mass of the visible displaced objects. We show the potential exclusion for the used luminosity in Ref.~\cite{ATLAS:2017tny} (32.8 fb$^{-1}$) as the orange dashed line (light orange shading) and the projected limit for the HL-LHC with 3 ab$^{-1}$ as the orange dot-dashed line (no shading).
The results are reinterpreted from Ref.~\cite{Heisig:2024xbh}. The search probes the larger mass splittings that are not covered by the existing HSCP and disappearing-track constraints. Its HL-LHC projection substantially extends the reach towards the CDFO boundary and covers most of the viable parameter space.

When comparing the minimal setup to the thermal-mass setup, we find that the limits do not change significantly. This is because the lifetime of the mediator $\phi$ is mostly determined by the Yukawa coupling $\lambda$ that is needed to obtain the correct relic abundance, and this coupling does not change significantly between the two thermal treatments.

\section{Conclusions}
\label{sec:concl}

We have shown that conversion-driven freeze-out can simultaneously account for the observed dark matter abundance and baryon asymmetry in quark-philic flavored DM models. The mechanism relies on semi-efficient conversions between the DM states and a colored mediator which is relatively close in mass. As these processes fall out of equilibrium, their $CP$-violating part generates
baryon asymmetries in the mediator and SM sectors. 
Electroweak sphalerons partially convert the SM-sector baryon asymmetry into a lepton asymmetry before they decouple, while the mediator baryon asymmetry is eventually transferred back to the SM sector by its decay. The resulting net baryon asymmetry is the observed BAU. %

We have studied a minimal realization with two Majorana DM flavors coupled to two flavors of right-handed down-type quarks. The small Yukawa couplings required for conversion-driven freeze-out, $\bar\lambda\sim10^{-7}$--$10^{-6}$, naturally lead to long-lived mediators and render conventional direct, indirect, flavor, and EDM probes ineffective. Nevertheless, a sizable baryon asymmetry can be generated through the resonant enhancement of the thermal $CP$ asymmetry associated with the small splitting between the two DM flavors. We find viable solutions throughout the conversion-driven freeze-out region, with DM masses ranging from a few hundred GeV up to about  $1.2\,\mathrm{TeV}$ 
and DM--mediator mass splittings  up to about $20\,\mathrm{GeV}$. The strong QCD interactions of the colored mediator, including bound-state effects, substantially enlarge the viable mass range compared with the leptophilic realization, while the impact of a
potentially
sizable Higgs-portal coupling remains mild.

The comparison of the minimal and thermal-mass setups shows that this regime lies in a technically challenging region where thermal corrections affect both the conversion dynamics and the $CP$-violating source. While our approximate treatments lead to the same qualitative picture, they can noticeably change the required $CP$ asymmetry and motivate a more complete and self-consistent finite-temperature calculation. Such a treatment may also sharpen, and potentially extend, the range of parameter space for which the mechanism can be assessed reliably.

The predicted mediator lifetimes can reach decay lengths of order meters, leading to complementary signatures in searches for heavy stable charged particles, disappearing tracks, and displaced vertices. Existing LHC searches already exclude small parts of the parameter space, particularly towards small mass splittings and large lifetimes. However, a dedicated displaced-vertex search for soft decay products -- proposed in the context of conversion-driven freeze-out -- can probe the larger-mass-splitting region that is not covered by the existing searches. Its HL-LHC projection covers most of the cosmologically viable region, providing a promising experimental test of the scenario.

While we have focused on scalar mediators coupled to right-handed SM fermions, the mechanism is not intrinsically tied to this minimal choice. It is expected to extend to models involving mediators coupled to $\mathrm{SU}(2)_L$ doublets, although the additional electroweak components, interactions, and chemical-equilibrium conditions require a dedicated analysis. Such extensions may also be interesting in leptophilic constructions with possible connections to neutrino-mass model building.

\section*{Acknowledgements}

We thank Michael Kr\"amer for valuable discussions and Andre Lessa for valuable help with the reinterpretation of the HL-LHC projections. This research was supported by the Deutsche Forschungsgemeinschaft (DFG, German Research Foundation) under grant 396021762 -- TRR~257.

\begin{appendix}

\section{Thermal corrections}  
\label{app:thermalcorrections}

In this appendix we illustrate the thermal corrections for $CP$-conserving processes that were adopted in the two setups presented in this work.\footnote{As for $CP$ violation in decay rates, we adapted the results from refs. \cite{Frossard:2012pc,Hambye:2016sby}.}


During the time relevant for our scenario (until around sphaleron decoupling, $T\gtrsim 130\,{\rm GeV}$), SM fermions and gauge bosons are in a regime of high energy-momentum $E\sim T$, $p\sim T$, $T\gg m_0=0$ (with $m_0$ the bare mass).
In this regime, the fermions closely resemble particles in vacuum with mass $\sqrt{2}\,m_f^{{\rm th}}$,
where $m_f^{{\rm th}}$ is the thermal mass~\cite{Klimov:1981ka,Weldon:1982bn,Pisarski:1988vb,Weldon:1989ys,Kiessig:2011fw}.
As for the gauge bosons, they propagate as transverse (physical) modes with thermal mass $m_g^{{\rm th}}=m_{D}/\sqrt{2}$ of order $gT$~\cite{Pisarski:1988vb,Pisarski:1988vb-gauge}, with $g$ the relevant gauge coupling and $m_{D}$ the Debye mass.
Therefore, we approximate the fields as free massive particles, with mass corresponding to the high momentum limit of the thermal mass
(see also refs. \cite{Landsman:1986uw,Kraemmer:1994az,Blaizot:2001nr,Bellac:2011kqa}):
\begin{align}
m_q^2(T) &=2\,m_q^{{\rm th}\,2}(T) =\frac{2}{6}\,g^2_s\,T^2,\\[1ex]
m^2_g(T) &= \frac{1}{2}\, m_{g,D}^2(T) 
=\frac{1}{2}\left(1+\frac{1}{6}N_f(T)\right)\,g^2_s\,T^2
\end{align}
where $m_q^{\rm th}(T)$ and $m_{g,D}$ denote respectively the quarks thermal mass and the gluon Debye mass.
As for the colored scalar field, we use \cite{Giudice:2003jh}
\begin{align}
 M_\Phi^2(T)
& =
 M_\Phi^2 + m_{\Phi,\mathrm{th}}^2(T)=M_\Phi^2 +\frac{2}{3}\,g^2_s\,T^2.
\end{align}

In our first setup with minimal thermal corrections, we use gluon and quarks thermal masses to delimitate the phase space of decay rates and cross sections.
The same masses
regulate the soft divergences of the scattering processes. 

In our second setup we additionally modify the particle propagators to estimate the $CP$-conserving part of the scattering processes.
In particular, we use the scalar propagator computed in real time formalism at one loop.

For a scalar particle, the propagator in momentum space
can be written in matrix form as~\cite{Niemi:1983nf,Giudice:2003jh}
\begin{equation}
\begin{aligned}
D(K)& =\begin{pmatrix} D^{11}(K) & e^{\beta \omega/2} D^-(K) \\ e^{-\beta \omega/2} D^+(K) & D^{22}(K) \end{pmatrix}
= \nonumber \\ &=
U(T,K)
\begin{pmatrix}
\bar{D}(K)& 0\\
0 & \bar{D}^*(K)
\end{pmatrix}U(T,K)\, ,
\end{aligned}
\end{equation}
The matrix $U$ is given by
\begin{equation}
  U(k) = \begin{pmatrix} \sqrt{1+n_B(\omega )} & \sqrt{n_B(\omega )} \\
                          \sqrt{n_B(\omega )}   & \sqrt{1+n_B(\omega )} \end{pmatrix}
\end{equation}
where $\omega= K^\alpha u_\alpha$ and $k= \big[ \big( K^\alpha u_\alpha \big)^2-K^2 \big]^{1/2}$ correspond to the Lorentz-invariant energy and three-momentum in the bath with four-velocity $u^\alpha$ ($u^\alpha u_\alpha =1$)~\cite{Weldon:1982aq}, and
\begin{equation}
    n_B(\omega )= \frac{1}{e^{|\omega|/T}-1}\, .
\end{equation}
$D^\pm$ are the cut propagators,
$\bar{D}(K)$ is the resummed propagator
\begin{equation}
D(K)=\frac{i}{K^2-m_{B,0}^2-\Sigma_B(K)+i\epsilon}\, ,
\end{equation}
where $m_B$ is the bare boson mass and 
$\Sigma_B(K)$ is the finite-temperature self-energy, which at one loop can be identified as the thermal mass $m_{B,0}^2+{\rm Re}\,\Sigma_B(K)=m_B^2(T)$.
Hence, for the propagator of the scalar field $\phi$ we use the expression
\begin{equation}
\begin{aligned}
  & D_{11}(K) = \bigl(D_{22}(K)\bigr)^* =
 \\ &
  = \frac{i}{K^2 -  M_\Phi^2(T)^2 + i\eta}
    + 2\pi n(\omega)\,\delta(K^2 -  M_\Phi^2(T)^2)
    \end{aligned}
\end{equation}

The (one-loop) resummed propagators for fermion fields can
be written as \cite{Kobes:1984vb,Giudice:2003jh}
\begin{align}
S(K)&=\left(
\begin{array}{cc}
S^{11}(K)& e^{\beta \omega/2} S^-(K)\\
e^{\beta \omega/2} S^+(K) & S^{22}(K)
\end{array}\right)=
\nonumber \\ & =
M\left(T,K\right)
\left(
\begin{array}{cc}
\bar{S}(K)& 0\\
0 & \bar{S}^*(K)
\end{array}\right)M\left(T,K\right)\, ,
\end{align}
where
\begin{equation}
M\left(T,K\right)=\left(
\begin{array}{cc}
\cos\phi_K & -\sin\phi_K \\
\sin\phi_K     & \cos\phi_K
\end{array}\right)
\end{equation}
with
\begin{equation}
n_F(\omega)\equiv \frac{1}{e^{|\omega|/T}+1}.
\end{equation}
and
\begin{equation}
\begin{aligned}
& \cos\phi_K=\left[\theta(\omega)-\theta(-\omega)\right]
\sqrt{1-n_F(\omega)}\, , \\
&  \sin\phi_K=
\sqrt{n_F(\omega)} \, . 
\end{aligned}
\end{equation}
$\bar{S}(K)$ is the resummed propagator
\begin{equation}
\bar{S}(K)=\frac{i}{\gamma^\mu K_\mu -m_0
-\Sigma_F(K)+i\epsilon}\, ,
\end{equation}
where $m_0$ is the fermion bare mass 
%
%
%
The self-energy $\Sigma_F(K)$ appearing in the denominator of the resummed propagator is of the form 
\cite{Weldon:1982bn}
\begin{align}
    & \Sigma(K)=-a\slashed{K}-b\slashed{u}
\end{align}
where 
$a,b$ are Lorentz-invariant functions, which can depend on the two Lorentz scalars $\omega, k$.
The poles in the propagator
occur when $\omega$ and $k$ are such as to produce a zero in the
denominator. In our regime of high momentum,
we approximate the dispersion relation
to that of a free particle,
$\omega=[2\,m^{\rm th}_f(T)^2+k^2]^{1/2}$, 
acquiring a
thermal mass $2\, m^{{\rm th}\,2}_f =2\, g^2 T^2 C(R)/8$, with $g$ the relevant gauge coupling and
$C(R)$ the quadratic Casimir invariant of the representation.
Hence, we use the approximate propagator
\begin{align}
S^{11}(K) & \approx  \slashed{K}
\bigg[\frac{i }{K^2-2\,m_q^{\rm th}(T)^2  + 
i\epsilon} 
\nonumber \\ & \hspace{1cm}
- 2\pi n_F(\omega)\delta \left(K^2-2\,m_F(T)^2
\right) \bigg] 
\end{align}

\end{appendix}

\bibliographystyle{bjstyle}
\bibliography{bib}

\end{document}